\documentclass[%
 amsmath,amssymb,
 aps,
]{revtex4-2}
\usepackage{amsmath}
\usepackage{float}
\usepackage{wrapfig}
\usepackage{graphicx}
\usepackage{dcolumn}
\usepackage[linesnumbered,ruled,vlined]{algorithm2e}
\usepackage{algpseudocode}
\usepackage{epstopdf}
\usepackage{cases}
\usepackage{bm}

\begin{document}

\preprint{APS/123-QED}

\title{Unsteady granular chute flows at high inertial numbers}

\author{Satyabrata Patro}
\affiliation{
 Department of Chemical Engineering, Indian Institute of Technology Kanpur, Uttar Pradesh, 208016, India
}
\author{Sumit Kumar}
\affiliation{Dr. Reddy's Laboratory, Hyderabad, Telangana, 500090, India}
\author{Anubhav Majumdar}
\affiliation{Oracle, Bengaluru, Karnataka, 560076, India}
\author{Anurag Tripathi}%
\email{anuragt@iitk.ac.in}
\affiliation{
 Department of Chemical Engineering, Indian Institute of Technology Kanpur, Uttar Pradesh, 208016, India
}

\date{\today}

\begin{abstract}
We study the time-dependent flow behavior of gravity-driven free surface granular flows using the discrete element method and continuum modeling. Discrete element method (DEM) simulations of slightly polydisperse disks flowing over a periodic chute with a bumpy base are performed. A simple numerical solution based on a continuum approach with the inertial number based $\mu-I$ rheology has been proposed to predict the flow dynamics. The results of the continuum model are compared with the DEM simulation results for a wide range of chute inclinations. Solutions for the constitutive model described by the popular JFP model as well as the recently proposed modified rheological model using a non-monotonic variation of $\mu-I$ are obtained. 
Our results demonstrate that the popular JFP model reliably predicts the flow at low to moderate inclination angles (i.e. for $I\lesssim0.5$). However, it fails to predict the flow properties at high inclinations. The modified rheological model, on the other hand, is very well able to predict the time-averaged flow properties for all the inclination angles considered in this study. Accounting for the presence of the slip velocity, layer dilation, and stress anisotropy are found to be crucial for accurate predictions of transient flows at high inertial numbers (i.e. for $I\gtrsim1$).
\end{abstract}

\maketitle
\section{\label{sec:Introduction}Introduction}
The rheology of granular materials has been an active research topic for the last few decades due to its wide occurrence in geophysical as well as industrial situations. A number of experimental \cite{pouliquen1999scaling, midi2004dense, jop2005crucial, holyoake2012high, pouliquen1999shape, heyman2017experimental, thompson2007granular, lacaze2008planar,lube2004axisymmetric,lajeunesse2004spreading, zhou2017experiments, khakhar1997radial, annular_shear_cell_2009comparisons, azanza1999experimental, yang2001simulation,orpe2007rheology} as well as simulation studies using discrete element method (DEM) \cite{silbert2001granular, midi2004dense, da2005rheophysics, pouliquen2006flow, baran_2006, borzsonyi2009patterns, tripathi2011rheology, kumaran2013, brodu_delannay_valance_richard_2015, mandal2016study, ralaiarisoa2017high, mandal2018study, bhateja2018, bhateja2020, patro2021rheology, debnath2022different} have been utilized to explore the rheology of granular materials. A detailed review of granular flow rheology in different configurations can be found in \cite{forterre2008flows, andereotti_forterre_pouliquen_2013, jop2015rheological}. These studies have shown that the granular flow between the two limiting cases of quasistatic, slow flows, and rapid, dilute flows is controlled by a non-dimensional inertial number $I$ which depends on the local shear rate and pressure in addition to the particle size and density. This intermediate dense flow regime has been studied in a variety of configurations. The inertial number-based rheological description for this dense flow regime has been confirmed in chute flows \cite{silbert2001granular, midi2004dense, jop2006constitutivenature, tripathi2011rheology, patro2021rheology}, plane shear flows \cite{da2005rheophysics,midi2004dense,mandal2016study,mandal2018study}, annular shear cell \cite{annular_shear_cell_2009comparisons}, granular collapse \cite{lube2004axisymmetric,lacaze2008planar, lagree_staron_popinet_2011,lajeunesse2004spreading}, planar silos \cite{bhateja2018,bhateja2020}, heap flows \cite{jop2005crucial} and rotating cylinders \cite{renouf2005numerical,orpe2007rheology}. Both experiments as well as simulations confirm that the ratio of the shear stress to the pressure depends on the inertial number in this regime.
The most popular model for the inertial number-based rheology is the JFP model \cite{jop2006constitutivenature}. 

According to this model, the effective friction coefficient $\mu$ varies in a nonlinear fashion with $I$. Starting from a minimum effective friction coefficient $\mu_s$ at $I\sim0$, $\mu$ increases with $I$. The inertial number-based JFP model \citep{jop2006constitutivenature} has been able to capture the flow behavior of granular materials in experiments \citep{forterre2003long,midi2004dense,jop2005crucial} as well as simulations \citep{da2005rheophysics,silbert2001granular,mandal2017sidewall} in the dense flow regime for inertial number $I\leq0.6$. In absence of simulation and experimental data at large $I$ values, the model assumes that the effective friction coefficient $\mu$ at large inertial numbers becomes constant. In addition, the solids fraction also decreases with the inertial number. Accounting for the solids fraction variation in the continuum simulations requires the incorporation of the compressibility effects. In order to use the commonly employed approach for incompressible fluid flows, most studies ignore the density variation in the continuum simulation of granular flows.
The incompressible JFP model has been used to predict the steady-state flow properties in chute flows \cite{jop2006constitutivenature, kamrin2010nonlinear, lagree_staron_popinet_2011, holyoake2012high,barker_gray_2017, barker2021OpenFoam,lin2020continuum,chauchat2014three}, vertical chute \cite{chauchat2014three}, steady plane shear \cite{lin2020continuum} as well as flow through an annular shear cell \cite{kamrin2010nonlinear,henann2013predictive}. The incompressible $\mu-I$ model or its variants have also been used to predict the transient flow of granular materials flowing over a chute \cite{barker_2021_gray,barker2015well,dunatunga2015continuum,lin2020continuum,franci20193d,parez2016unsteady,barker_gray_2017}, material discharging from granular silos \cite{staron-2012, staron2014continuum, kamrin2010nonlinear,dunatunga2015continuum}, flow during column collapse \cite{lagree_staron_popinet_2011,dunatunga2015continuum,lin2020continuum,franci20193d,rauter2021compressible}, in addition to rotating drum \cite{barker_2021_gray}, heap flows \cite{chauchat2014three}.

The time-dependent response of granular media in bounded heap flow has been studied using DEM simulations and experiments \cite{boundedheapflow2017}. Efforts have also been made to capture the experimentally observed transient behavior of granular slides in air as well as water \citep{submergedlandslide2019}. The depth-averaged model approach has also been utilized in the past to predict the flow behavior of granular materials over a chute \cite{gray2014depth,rocha2019self,delannay2017granular,mangeney2007numerical}. 
Granular flows also exhibit some non-local effects during slow flows such as shear banding, weak dependence of stress on the magnitude of shear rate, and dilation effects \cite{dsouzaNott2020nonlocal}. Some non-local continuum models have also been proposed to understand the flow behavior of granular materials \cite{pouliquen2009non,henann2013predictive,kamrin2012nonlocal}. Recently \citet{debnath2023comparison} have solved and compared different compressible $\mu-I$ \cite{barker2017wellposedrheology,schaeffer2019constitutive} as well as non-local rheological models \cite{henann2013predictive,dsouzaNott2020nonlocal} with the DEM simulations of flow in a vertical chute. These models are found to have limited success in predicting the flow properties. A review of the non-local modeling for granular flows is presented by \cite{kamrin2019non}.
Concerns about studies dealing with time-dependent continuum simulations have been raised in the past. \citet{barker2015well} have shown that the incompressible JFP model is well-posed only for a narrow range of inertial numbers; for very low and relatively high values of inertial numbers, it is found to be ill-posed. A wider range of well-posed regions can be obtained by partial regularization of the $\mu-I$ model by deriving a new functional form of the $\mu-I$ model so that the granular material no longer has yield stress \cite{barker_gray_2017}. 
The compressible rheology has been implemented to understand the behavior of subaqueous granular collapse \cite{rauter2021compressible}.
Few efforts have also been made to introduce bulk compressibility effects \cite{heyman2017compressibility, barker2017wellposedrheology, schaeffer2019constitutive} to regularize the $\mu-I$ model. Recently \citet{barker_2021_gray} have implemented the regularized well-posed $\mu-I$ rheology \cite{barker_gray_2017} intercoupled with segregation in a continuum framework to capture the qualitative behavior of flow evolution and segregation of multi-component mixtures in an inclined plane and rotating tumbler.
\vspace{1cm}
 
 \begin{figure*}[hbtp]
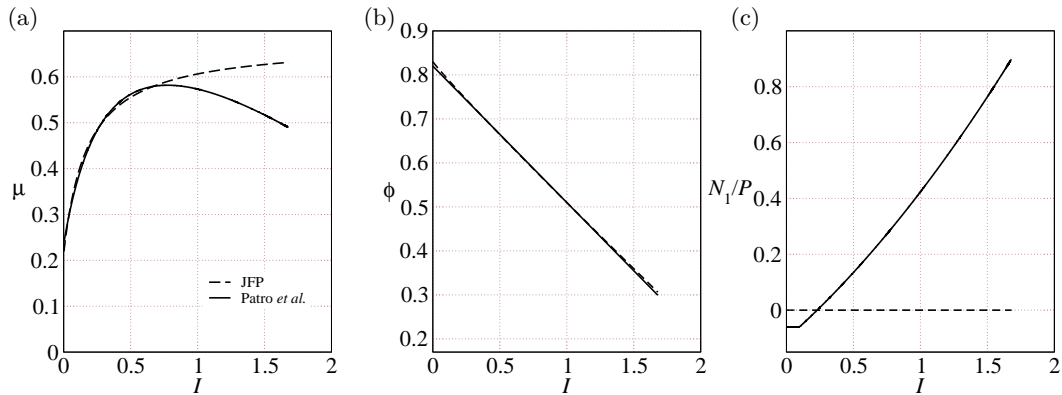

\includegraphics[scale=0.33]{Fig_1a.eps}\put (-125,140) {(a)}
\hspace{0.4cm}
\includegraphics[scale=0.33]{Fig_1b.eps}\put (-130,140) {(b)}
\includegraphics[scale=0.33]{Fig_1c.eps}\put (-125,140) {(c)}
\caption{Variation of the (a) Effective friction coefficient $\mu$ with the inertial number $I$ (b) Solids fraction $\phi$ with $I$ and (c) normal stress difference to the pressure ratio $N_1/P$ with $I$ using the model parameters from \citet{patro2021rheology}. Solid lines represent the variation according to the modified rheology. The dashed line represents the fitted line using the JFP model.}
\label{Figure_1}
\end{figure*}

All of the aforementioned studies using inertial number rheology employ the assumption that the effective friction coefficient becomes constant at high inertial numbers. \citet{mandal2016study,mandal2018study}, however, in their simulation study of the flow of dumbbells in plane shear flows showed that the saturating behavior of the effective friction coefficient $\mu$ at higher inertial number $I$ used in such studies is not correct. Instead, $\mu$ is found to decrease with $I$ after achieving a maximum. However, the authors were unable to observe similar behavior in the case of chute flows. A recent study by \citet{patro2021rheology} has shown that the non-monotonic variation of the effective friction coefficient $\mu-I$ with the inertial number $I$ is observed in the case of chute flows as well. Figure~\ref{Figure_1}(a) shows the variation of the effective friction coefficient with the inertial number (solid line) as suggested by the modified rheology of \citet{patro2021rheology}. The broken line shows the variation according to the JFP model.
Figure~\ref{Figure_1}(b) shows the variation of the solids fraction $\phi$ with the inertial number $I$ for the two models. Figure~\ref{Figure_1}(c) shows the variation of the ratio of the normal stress difference to the pressure ratio $N_1/P$ with the inertial number $I$. Note that the JFP model does not account for the presence of normal stress difference and hence $N_1=0$ at all inertial numbers. Non-zero normal stress differences in granular flows, however, have been observed by other researchers \cite{goldhirsch1996origin,alam2003first,saha2016normal,srivastava2021viscometric,santos2022}. By accounting for the normal stress difference law in the modified rheological model, \citet{patro2021rheology} also confirmed that two different flow states at the same inclination angle of the chute are not possible despite the non-monotonic variation of the $\mu-I$ with the $I$. The authors also showed that the modified rheological description coupled with momentum balance equations is able to predict various flow properties of interest at steady state for chute flow of disks at different inclination angles. These predictions have been found to be in good agreement with DEM simulation results for periodic chute flows even at high inertial numbers. However, predictions for time-dependent properties for high inertial number granular flows have not been compared with DEM simulations to the best of our knowledge.

In this study, we focus on unsteady granular flows down an inclined surface in a periodic chute flow configuration spanning a large range of inclination angles to cover a wide range of inertial numbers. \citet{parez2016unsteady} have obtained analytical expressions by solving the momentum balance equation coupled with a linear $\mu-I$ relation for such a system. Extending this analytical approach to the non-linear $\mu-I$ rheological model is much more complex and mathematically challenging. Hence, we solve the resulting equation numerically to predict the time-dependent flow properties. The predictions obtained from the numerical solution are compared with the DEM simulation results for the first time for inertial numbers close to or greater than unity. Our results show that using the JFP model at high inertial number flows leads to significant overprediction of the average velocity along with underprediction of solids fraction.

The organization of the paper is as follows. The DEM simulation methodology is briefly mentioned in Section II. The theoretical formulation as well as the numerical technique used to solve the resulting equation is described in Section III. Results obtained from the numerical solution of the momentum balance equation are compared with the DEM simulations in Section IV. Discussion about some important observations is reported in Section V. Summary and future work are presented in Section VI.

\section{\label{sec:simulation_methodology}Simulation methodology}

\begin{figure*}
	\centering
		\includegraphics[scale= 0.15]{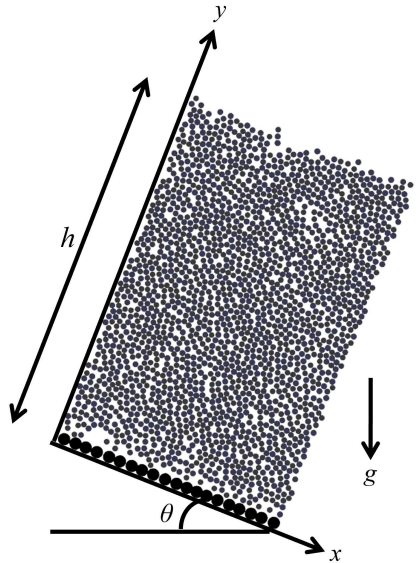}
	\caption {Snapshot of simulation of a granular layer flowing under the influence of gravity at an inclination $\theta=32^\circ$ from the horizontal at any instant. Black-filled circles represent the static particles that form the bumpy chute base and grey-filled circles represent the flowing discs. }
	\label{Figure_2}
\end{figure*}

The discrete element method (DEM) technique is used to simulate slightly poly-disperse ($\pm 5 \%$ polydispersity), inelastic, frictional disks flowing over a rough and bumpy inclined surface. A schematic of the simulation setup is shown in Fig.~\ref{Figure_2}. The length of the simulation box is $40d$ where $d$ is the mean diameter of the discs. To mimic an infinite length of the chute, a periodic boundary condition is imposed in the flowing ($x$) direction. The base of the simulation domain consists of static bumpy particles of size $2d$ to reduce the slip at the base.
The contact force between the discs is modeled using the linear spring and dashpot model as in the L2 model of \citet{silbert2001granular}. 
The coefficient of static friction $\mu$ is chosen as 0.5 and two different values of the normal restitution coefficient ($e_n=0.5~\&~0.1)$ are considered. 
The initial arrangement of the particles is done on a square lattice with a finite spacing between the surface of particles and they are given small random initial velocities. In order to simulate the flow of particles from rest, the particles are allowed to settle under the influence of gravity until the average kinetic energy of the particles in the layer becomes less than $10^{-6}mgd$. The height of the settled layer of $N=2000$ particles following this protocol is $h\sim50d$. 
At time $t=0$, the direction of the gravity is changed to the desired inclination angle $\theta$. The flow is allowed to evolve until the average kinetic energy of the system of particles becomes constant, indicating that the system has achieved a steady state. 
To compute the flow properties, the simulation domain is divided into a number of strips in the $y$ direction. The thickness of each strip is equal to the mean particle diameter $d$ so that the reported properties are averaged over the strip area $A=L_x\times d$. The properties reported at any instant $t$ are averaged over 60 snapshots having an interval of $0.1$ time unit between successive snapshots and thus represent the average property over the last 6 time units. The partial contribution of the particle's area in each strip is accounted for while calculating various flow profiles such as velocity, solids fraction, stresses, etc. More details are available in \citet{patro2021rheology}.

\section{\label{sec:theory}Theory}

Consider a fully developed granular flow over a surface inclined at an angle $\theta$ under the influence of gravity. Assuming a unidirectional flow in the $x$ direction, the momentum balance equation in $x$ and $y$ directions simplifies to
    \begin{equation}
    \rho_b\frac{\partial v_x}{\partial t} =  - \frac{\partial\tau_{yx}}{\partial y}+\rho_b g \sin\theta,
    \label{x_momentum}
    \end{equation}
    \begin{equation}
    0=-\frac{\partial{\sigma_{yy}}}{\partial{y}}-\rho_b{g}{\cos\theta},
    \label{sigma_yy_momentum_balance_y_1st}
    \end{equation}
where $v_x$ is the velocity along the flow direction, $\theta$ is the inclination angle, $\tau_{yx}$ is the shear stress, $g$ is the gravitational acceleration, and $\rho_b=\phi(y)\rho_p$ is the bulk density of the medium with $\rho_p$ being the density of the particle and $\phi(y)$ is the local solids fraction at any $y$.
Integrating Equation~(\ref{sigma_yy_momentum_balance_y_1st}) with $y$, we get 
    \begin{equation}
    \sigma_{yy}=\rho_p{g}{\cos\theta}\int_{y}^{h}\phi(y)dy.
    \label{sigma_yy_momentum_balance_y_2nd}
    \end{equation}
Assuming the variation of $\phi(y)$ along $y$ at any instant to be small, we approximate the integral in Eq.~(\ref{sigma_yy_momentum_balance_y_2nd}) as $\int_{y}^{h}\phi(y)dy=\phi_{avg}(h-y)$, so that the expression for $\sigma_{yy}$ simplifies to
    \begin{equation}
    \sigma_{yy}=\rho_p{g}{\cos\theta}\phi_{avg}(h-y).
    \label{sigma_yy_momentum_balance_y_3rd}
    \end{equation}
According to the $\mu-I$ rheology, the flow behavior depends upon the non-dimensional inertial number $I$ defined as
\begin{equation}
I=\frac{\dot{|\gamma|}d}{\sqrt{P/\rho_p}}.
\label{inertialnumber}
\end{equation}
For the case of unidirectional chute flow, the second invariant of the strain rate tensor $\dot{|\gamma|}$ equals the shear rate $dv_x/dy$, i.e., $\dot{|\gamma|}=dv_x/dy$. The effective friction coefficient $\mu(I)$ is defined as the ratio of the second invariant of the stress tensor $|\tau_{yx}|$ to the pressure $P$, i.e.,
    \begin{equation}
    \mu(I)=\frac{|\tau_{yx}|}{P}.
    \label{mu_I_general_equation}
    \end{equation} 
The JFP model uses the following form to relate the effective friction coefficient $\mu(I)$ with $I$
\begin{equation}
\mu(I)=\mu_{s}^{'}+\frac{\mu_{m}^{'}-\mu_{s}^{'}}{1+I_{0}^{'}/I},
\label{mu_i_JFP}
\end{equation}
with $\mu_{s}^{'}$, $\mu_{m}^{'}$ and $I_{0}^{'}$ being the model parameters.
Our recent study~\cite{patro2021rheology} shows that a more appropriate variation of $\mu(I)$ is given by 
    \begin{equation}
    \mu(I)=\mu_{s}+\frac{c_{1}-c_{2}I}{1+I_{0}/I}.
    \label{mu_i_MK}
    \end{equation}
The proposed rheology is complemented with an empirical relation to describe the variation of solid fraction ($\phi$) with inertial number $I$ as 
    \begin{equation}
    \phi=\phi_{max}-aI^\alpha
    \label{phi_nonlinear}
    \end{equation}
    where $\phi_{max}$, $a$ and $\alpha$ are the model parameters \cite{patro2021rheology}.
In addition, we find that the role of normal stress difference becomes important at high inclinations. This presence of the normal stress difference is accounted by a normal stress difference law in the rheology by proposing the ratio of the first normal difference $N_1=\sigma_{xx}-\sigma_{yy}$ to the pressure $P=(\sigma_{xx}+\sigma_{yy})/2$ as a function of inertial number $I$, i.e.,
    \begin{equation}
    \frac{N_{1}}{P}=f(I).
    \label{n1byp_f(I)}
    \end{equation}
Using Eqs.~(\ref{sigma_yy_momentum_balance_y_3rd}), (\ref{mu_I_general_equation}) and (\ref{n1byp_f(I)}) we get the expression for pressure as
    \begin{equation}
    P = \frac{2\phi_{avg}\rho_p g \cos\theta (h-y)}{2-f(I)}.
    \label{P_equation}
    \end{equation}
As expected, in absence of normal stress difference, the expression for pressure reduces to a hydrostatic head. Recent study by \citet{patro2021rheology} shows that $f(I)$ remains constant up to $I\leq I^*$ and varies quadratically with inertial number $I$ for $I>I^*(=0.1)$, i.e.,

    \begin{subnumcases} 
    {f(I) =}
    k & for $I\leq0.1$,\\ \label{constant_fI_equation}
    AI^2+BI+C & for $I>0.1$.\label{quadratic_P_equation}
    \end{subnumcases}
    Using the expression of pressure $P$ from Eq.~(\ref{P_equation}) in Eq.~(\ref{inertialnumber}), we get,
    \begin{equation}
    I = \frac{\dot{|\gamma|}d\sqrt{2-f(I)}}{\sqrt{2g\cos\theta (h-y)\phi_{avg}}}.
    \label{I_equation}
    \end{equation}
    Equation~(\ref{I_equation}) can be rearranged to get the expression for $I$ as follows:
    \begin{subnumcases} {I=}
    \frac{- B_0 + \sqrt{B_0^2 - 4A_0C_0}}{2A_0} & for $I>0.1$\label{solve_Ia}\\
    \frac{\dot{|\gamma|}d\sqrt{2-k}}{\sqrt{2g\cos\theta (h-y)\phi_{avg}}} & for $I\leq0.1$
    \label{solve_Ib}
    \end{subnumcases}
with $A_0=2g\cos\theta (h-y) \phi_{avg} + A\dot{|\gamma|}^2d^2$, $B_0=B\dot{|\gamma|}^2d^2$ and $C_0=(C-2)\dot{|\gamma|}^2d^2$.
The calculated inertial number is used in the empirical form relating the effective friction coefficient $\mu(I)$ with the inertial number $I$. Eqs.~(\ref{mu_i_JFP}) and (\ref{mu_i_MK}) show two such empirical forms that can be utilized to obtain the time-dependent properties of the flow .
Using Eqs.~(\ref{mu_I_general_equation}) and (\ref{P_equation}), we get
    \begin{equation}
    \tau_{yx} = \mu(I)\left(\frac{2\phi_{avg}\rho_p g \cos\theta (h-y)}{2-f(I)}\right).
    \label{tau_equation}
    \end{equation}
    Differentiating Eq.~(\ref{tau_equation}) with respect to $y$ and substituting $\frac{\partial\tau_{yx}}{\partial y}$ in Eq.~(\ref{x_momentum}), we get
    \begin{equation}
    \phi(I)\frac{\partial v_{x}}{\partial t} = \phi(I)g \sin\theta -  \frac{\partial}{\partial y}\left[\mu(I)\phi_{avg}\left(\frac{2\rho_p g \cos\theta (h-y)}{2-f(I)}\right)\right]
    \label{x_momentum_detail}.
    \end{equation}
    
In writing Eq.~(\ref{x_momentum_detail}), we account for the variation of $\phi$ with $y$ in the $x$ momentum balance equation. However, in the calculation of $\sigma_{yy}$, the variation of $\phi$ with $y$ is ignored and $\phi(y)$ is replaced by the average value $\phi_{avg}$ (Eq.~(\ref{sigma_yy_momentum_balance_y_3rd})). We also solve Eq.~(\ref{x_momentum_detail}) by accounting for the variation of $\phi$ along $y$ in the calculation of $\sigma_{yy}$ using Eq.~(\ref{sigma_yy_momentum_balance_y_2nd}). We find that the results obtained are not altered significantly due to this more refined calculation of the $\sigma_{yy}$. Hence we use Eq.~(\ref{sigma_yy_momentum_balance_y_3rd}) for the calculation of $\sigma_{yy}$ in all the results presented in this work.  
In order to solve Equation~(\ref{x_momentum_detail}), we use the PDEPE solver in MATLAB along with the following initial and boundary conditions:
 
\begin{subequations}
\begin{align}
IC: v_{x}(y,0) & = 0,\label{IC}\\
BC\hspace{0.1cm}1: \tau_{yx}(h,t) & = 0,\label{BC1}\\
BC\hspace{0.1cm}2: v_{x}(0,t) & = v_{slip}(t).\label{BC2}
\end{align}
\end{subequations}
    
The initial condition (Eq.~(\ref{IC})) represents that the velocity across the layer at $t=0$ is zero and mimics the flow starting from rest. The first boundary condition (Eq.~(\ref{BC1})) corresponds to zero shear stress at the free surface $y=h$. The second boundary condition (Eq.~(\ref{BC2})) corresponds to a known slip velocity at the base. While a no-slip boundary condition seems to be appropriate for low inclinations, this boundary condition is needed to account for the sufficient slip observed at high inclinations \cite{patro2021rheology}.
  The general form of the partial differential equation used by $PDEPE$ solver in MATLAB is given as follows:
    \begin{equation}
    \small
    c\left(y,t,u,\frac{\partial u}{\partial y}\right)\frac{\partial u}{\partial t} = y^{-m}\frac{\partial}{\partial y}\left(y^mF\left(y,t,u,\frac{\partial u}{\partial y}\right)\right)+s\left(y,t,u,\frac{\partial u}{\partial y}\right).
    \label{general_pde}
    \end{equation}
Equation~(\ref{x_momentum_detail}) can be written in the above general form using with $u=v_x$ and $c=\phi(I)$, $m=0$, $s=\phi(I)g\sin\theta$ and $F=\mu(I)\phi(I)\left(\dfrac{2\rho_p g \cos\theta (h-y)}{2-f(I)}\right)$.

The PDEPE solver represents the derivatives numerically by discretizing the domain into finite spatial and temporal grids. The number of grids along the $y$ direction is chosen to be $N=50$ and the time step $\Delta t$ is chosen to be $0.1$. 
Equation~(\ref{phi_nonlinear}), which is referred to as the dilatancy law, dictates that the solids fraction $\phi$ of the medium decreases with an increase in the inertial number. With increasing velocity (and hence inertial number), the flowing granular layer dilates and leads to an increase in the layer thickness $h$ with time $t$. Our DEM simulation results show that this increase in the layer height becomes very significant at high inertial numbers. 
This significant dilation of the granular layer indicates that the compressibility effects become important at high inertial numbers and need to be accounted for by solving the continuity equation as well. Given our assumption of the unidirectional flow, the $y$-component of the velocity is ignored in our theoretical formulation. Due to this reason, accounting for the variation of $\phi$ with $t$ is not possible using the equation of continuity. We circumvent this problem by using the integral mass balance equation. Since our  
simulation method utilizes a periodic simulation box with a fixed mass of particles, we account for the height of the layer at any instant by equating the mass at time $t$ to the mass in the beginning of the simulation.
The height of the layer is $h_{t=0}=h_{min}$ at $t=0$ as it starts from the rest with the maximum solids fraction $\phi_{t=0}=\phi_{max}$ across the entire layer.
The mass per unit width of the flowing layer of the disks at any time instant $t$ is equal to $\int_{0}^{h(t)}\rho_b(t)dy$. Using $\rho_b(t)=\phi(t)\rho_p$ and equating the mass at any instant to the initial mass of the layer, we get $\int_{0}^{h(t)}\phi(t)dy=h_{min}\phi_{max}$. Since $\int_{0}^{h(t)}\phi(t)dy=h(t)\phi_{avg}(t)$, the height at any instant can be obtained using $h(t)=\dfrac{h_{min}\phi_{max}}{\phi_{avg}(t)} $ where $\phi_{avg}(t)$ the average solids fraction across the layer at any instant $t$. The detailed steps for computing the time-dependent properties of the flowing layer are given in Algorithm~\ref{ALGO}. 

\begin{algorithm}
\caption{Prediction of time-dependent flow properties}
\label{ALGO}
\SetAlgoLined
\DontPrintSemicolon
Initialize $h=h_{min}$, $\phi(y)=\phi_{avg}=\phi_{max}$, $\epsilon=10^{-5}$, temporal grid size $\Delta t=0.1$, number of spatial grid points in $y$ direction $N=50$.\;
Obtain discrete positions $y=[0:\Delta y:h]$ along the $y$ direction using a uniform spatial grid spacing $\Delta y=h/N$.\;
Use Equation (\ref{IC}) for initial condition $v_{x}(y)=0$ at $t=0$.\;
\While {$t < t_{final}$}{
Calculate the following properties at time $t$.\;
Velocity gradient $\dot{|\gamma|}=dv_{x}(y)/dy$.\;
Inertial number $I(y)$ using Eqs. (\ref{solve_Ia}) \& (\ref{solve_Ib}).\;
Solids fraction $\phi(y)$ using Eq. (\ref{phi_nonlinear}).\;
Pressure $P(y)$ using Eq. (\ref{P_equation}).\;
Shear stress $\tau_{yx}(y)$ using Eq. (\ref{tau_equation}).\;
Solve Equation (\ref{x_momentum_detail}) with IC, BC 1 and BC 2 (Eqs. (\ref{IC})-(\ref{BC2})) using $PDEPE$ to obtain velocity $v_{x, next}$ at $t+\Delta t$.\;
Calculate the following properties at time $t+\Delta t$.\;
Velocity gradient $dv_{x,next}(y)/dy$.\;
Inertial number $I_{next}(y)$ using Eqs. (\ref{solve_Ia}) \& (\ref{solve_Ib}).\;
Solids fraction $\phi_{next}(y)$ using Eq. (\ref{phi_nonlinear}).\;
Average solids fraction $\phi_{avg, next}$ by averaging the $\phi_{next}(y)$ across the bulk layer.\;
\eIf {$((||v_{x,next} - v_{x}||) > \epsilon$}{
Update the flowing layer thickness $h_{next}=h\phi_{avg}/\phi_{avg,next}$.\;
Recalculate discrete positions $y_{next}=[0:\Delta y_{next}:h_{next}]$ along the $y$ direction using a uniform spatial grid spacing $\Delta y_{next}=h_{next}/N$.\;
Assign $v_{x}(y)=v_{x,next}(y_{next})$, $\phi(y)=\phi_{next}(y_{next})$, $h=h_{next}$ and $\Delta y=h_{next}/50$.\;
Update time as $t=t+\Delta t$\;}
{Exit\;}
}
\end{algorithm}

\section{\label{sec:Results}Results}
In this section, we present results for the time-dependent flow of disks over an inclined surface using DEM simulations and compare them with the theoretical predictions. The results are reported in dimensionless form for two different restitution coefficients $e_n=0.5$ and $e_n=0.1$. The theoretical predictions are obtained by solving the momentum balance equations along with the inertial number-based JFP model popularly used for describing the behavior of dense granular flows. Due to the limitations of the JFP model to capture the behavior at high inertial numbers, the recently proposed MK model along with a normal stress difference law has been used to predict the flow properties for dense as well as dilute granular flows. We first compare the DEM simulation results with the analytical predictions of \citet{parez2016unsteady}. Since the analytical expressions derived by the authors assume a linear $\mu-I$ relation, we compare their theory only for $I\leq0.35$. Next, we compare the DEM results with the predictions obtained using two different rheological models
\subsection{Analytical and numerical predictions for linear $\mu-I$ model}
\begin{figure*}
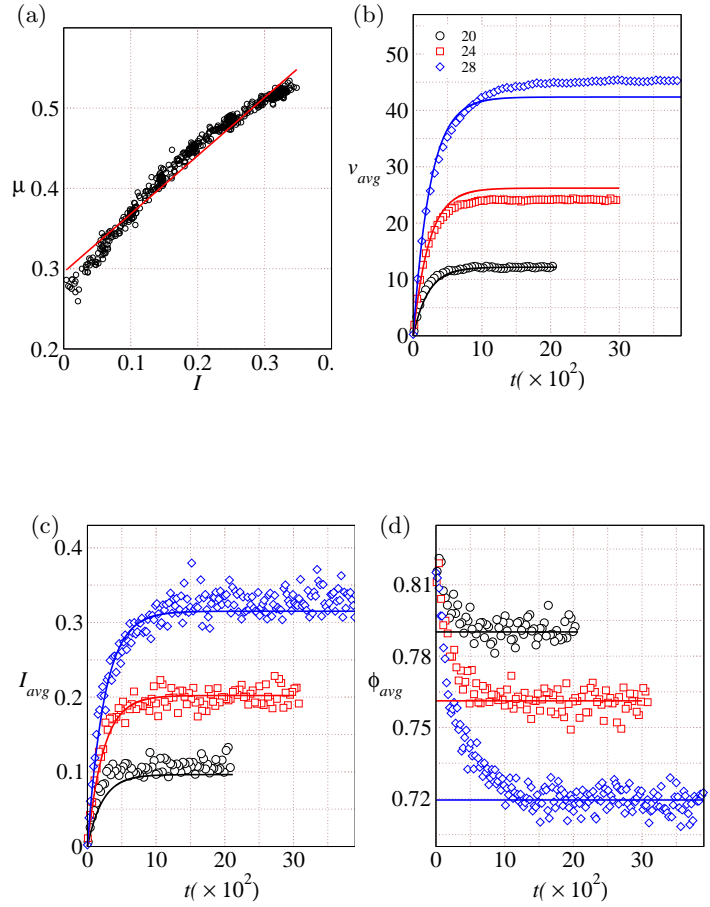

 	\includegraphics[scale=0.33]{Fig_3a.eps}
 	\put (-125,140) {(a)}
 	\includegraphics[scale=0.33]{Fig_3b.eps}
	\put (-125,140){(b)}
	\hspace{0.5cm}
	
	\vspace{1.6cm}
	\includegraphics[scale=0.33]{Fig_3c.eps}
	\put (-125,140){(c)}
	\includegraphics[scale=0.33]{Fig_3d.eps}
	\put (-125,140){(d)}
	\caption{(a) Variation of effective friction coefficient $\mu$ with inertial number $I$ for $e_n=0.5$. Black circles represent the DEM data up to inclination $\theta=28^\circ$. The red solid line represents the fitted line of the form $\mu(I)=aI+b$.  Variation of the (b) average velocity of the layer $v_{avg}$, (c) average inertial number in the bulk $I_{bulk}$, and (d) average solids fraction in the bulk $\phi_{bulk}$ with time $t$ for $e_n=0.5$ at different inclinations. Symbols represent the DEM simulation data and the solid lines represent the analytical predictions by \citet{parez2016unsteady}.}
\label{Figure_3}
\end{figure*}

In this section, we compare the predictions of inertial number, velocity profile, and solids fraction using the analytical solution given by \citet{parez2016unsteady}. The authors predicted the flow properties for unsteady flows flowing down an inclined plane using an analytical approach. They considered a linear form of the $\mu-I$ and their results were in very excellent agreement with their DEM simulation results. Due to the weak dependence of solids fraction on the flow velocity reported in previous works \cite{forterre2008flows,silbert2001granular,midi2004dense,da2005rheophysics,shearflowofgranularmaterials2012}, the authors assumed a constant solids fraction in their analytical approach to derive the expression for time-dependent flow properties. The authors obtained a series solution for the shear rate and the velocity profile. The solution can be very well approximated by a single-term solution and ignoring other terms of the series since they are much smaller compared to the first term. This analytical solution for the time-dependent velocity profile was derived using a linear relation between effective friction coefficient $\mu(I)$ with the inertial number $I$, i.e, $\mu(I)=\tan\theta_r+bI$. The values of the model parameters $\tan\theta_r$ and $b$ are obtained by fitting a line to the $\mu-I$ data obtained from DEM simulations as shown in Fig.~\ref{Figure_3}(a). In order to ensure linear $\mu-I$ relation, only data up to $\theta\leq28^\circ$ are considered in Fig.~\ref{Figure_3}(a). The value of the slope and intercept of the fitted line are obtained to be $b=0.76$ and $\tan\theta_r=0.29$. The solid lines in Figs.~\ref{Figure_3}(b)-\ref{Figure_3}(d) show the flow properties such as the average velocity, the bulk inertial number and the bulk solids fraction predicted using the single-term analytical solution. 
We use the following expressions reported by \citet{parez2016unsteady} for velocity
\begin{equation}
    v(y,t) = \dfrac{2}{3}\dfrac{\sqrt{\phi_{s}g \cos{\theta}} (\tan{\theta}-\tan{\theta_r})}{bd}(h_s^{3/2}-y^{3/2})(1-e^{-t/T_1}) + h.o.t.
    \label{vx_parez}
\end{equation}
where 
$T_1 =0.5\sqrt{\phi_s}\dfrac{h_s^{3/2}}{bd\sqrt{g \cos{\theta}}}$. 

The inertial number is calculated using Eq.~(\ref{I_equation}) where the shear rate is obtained by numerically differentiating the  velocity predicted from the theory of \citet{parez2016unsteady} and the pressure is assumed to be equal to hydrostatic pressure, following the assumptions of the theory.
Note that \citet{parez2016unsteady} assume that the height of the flowing layer $h$ as well as solids fraction $\phi$ remains constant and does not change with time. Due to this reason the time variation of the average solids fraction is not captured by their theory. We use the steady-state height $h_s$ and solids fraction $\phi_s$ in Eq.~(\ref{vx_parez}) to calculate the predictions shown in Figure~\ref{Figure_3}. DEM simulation results are shown using symbols. The profiles predicted using their analytical solution are in reasonable agreement with the DEM data up to $I=0.3$. 
At earlier times, the difference between theory and simulations is observable due to the difference in the $\mu-I$ data and the linear fit used. Fitting a line only for data lying in the range of $I<0.1$ improves the analytical predictions for $\theta=20^\circ$. Similarly, a better match can be obtained for $\theta=24^\circ$ by using a linear fit only up to $I\sim0.2$.
At a higher inclination angle of $\theta=28^\circ$, the analytical predictions for $v_{avg}$ differ from simulation data not only for early times but also for later times. Increasing the inclination angle further leads to more deviation from the simulation data since the non-linear behavior of the data is not captured well using a linear fit for higher values of the inertial number observed at these high inclinations. In absence of any theoretical results for non-linear $\mu-I$ relation, we resort to the numerical solution of the momentum balance equations in the next sections. Before reporting the numerical results of momentum balance equations for the non-linear $\mu-I$ rheology, we compare our numerical predictions for the linear $\mu-I$ case with the analytical solutions of \citet{parez2016unsteady}.
\begin{figure*}[hbtp]
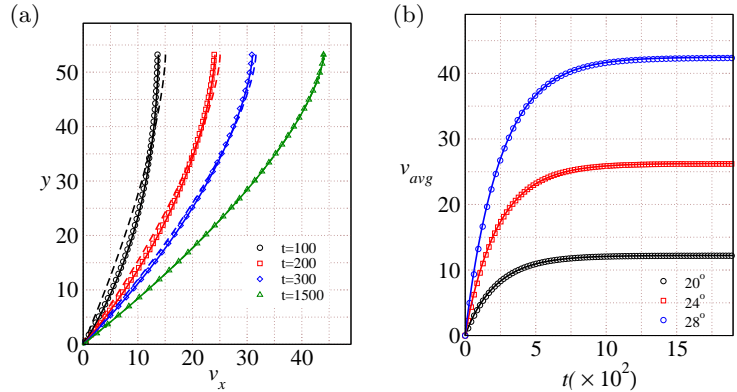

 	\includegraphics[scale=0.33]{Fig_4a.eps}
        \put (-130,140) {(a)} 
 	\hspace{0.5cm}
 		\includegraphics[scale=0.33]{Fig_4b.eps}
        \put (-130,140) {(b)}
 	    \caption{a) Comparison of the numerical solution (shown using symbols), analytical one-term solution (dashed line), and analytical three-term solution (solid lines) at different times for $\theta=24^\circ$. b) Comparison of the numerical solution (shown using symbols) and analytical one-term solution (solid lines) for different angles.}
      \label{Figure_4}
\end{figure*}

We first benchmark our numerical solutions using the rheological parameters reported in \citet{parez2016unsteady} and find a near-perfect match with their analytical solutions (not shown here). Next, we use the fitted parameter for linear $\mu-I$ relation from Fig.~\ref{Figure_3}(a) to obtain numerical solutions corresponding to the analytical solutions shown in Fig.~\ref{Figure_3}(b). Figure~\ref{Figure_4}(a) shows the results for velocity profile at different time instants for $\theta=24^\circ$. Figure~\ref{Figure_4}(b) shows the average velocity $v_{avg}$ variation with time for three different inclinations. The dashed line corresponds to the single-term solution while the solid line accounts for the first three terms of the infinite series solution. The slight difference between the single-term solution and the three-term solution is due to the fact that the neglected terms of the series solution remain comparable to the first term of the series at smaller times and become negligible only at large times. In the next sections, we use this numerical method with non-linear $\mu-I$ relation to obtain predictions for the flow properties with time. The excellent match between the numerical solutions (symbols) and analytical predictions (solid lines) confirms the accuracy  of the numerical solutions. 
\vspace{1.7cm}    

    \begin{figure*}[h!]
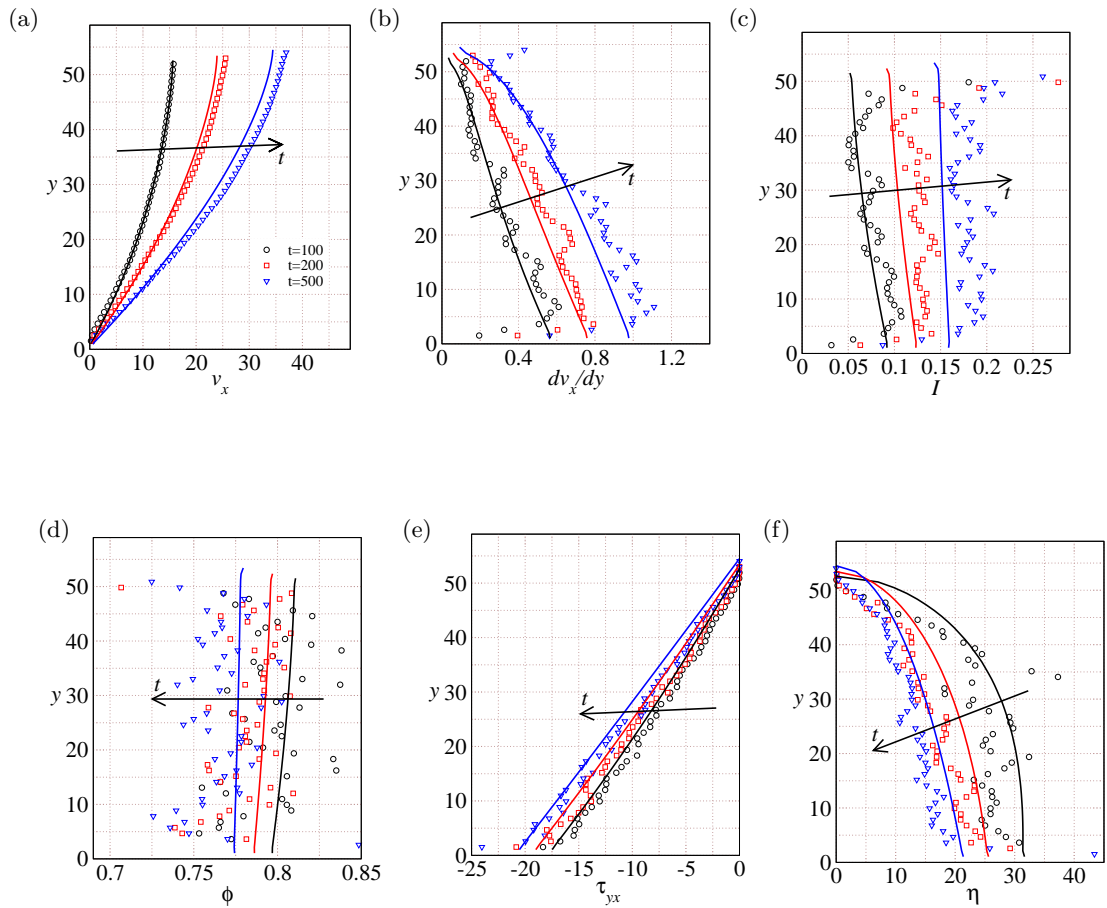

 	\includegraphics[scale=0.33]{Fig_5a.eps}\put (-130,140) {(a)}
 	\hspace{0.5cm}
	\includegraphics[scale=0.33]{Fig_5b.eps}\put (-130,140){(b)}
	\hspace{0.5cm}
	\includegraphics[scale=0.33]{Fig_5c.eps}\put (-130,140){(c)}
	\hspace{0.5cm}
	
	\vspace{1.6cm}
	\includegraphics[scale=0.33]{Fig_5d.eps}\put (-130,140){(d)}
 	\hspace{0.5cm}
	\includegraphics[scale=0.33]{Fig_5e.eps}\put (-130,140){(e)}
	\hspace{0.5cm}
	\includegraphics[scale=0.33]{Fig_5f.eps}\put (-130,140){(f)}
		\caption{Variation of the (a) velocity $v_x$, (b) shear rate $\dot{\gamma}$, (c) inertial number $I$, (d) solids fraction $\phi$, (e) shear stress $\tau_{yx}$ and (f) viscosity $\eta$ with distance $y$ from the base at different times for $\theta=24^\circ$. Symbols represent the DEM simulations data and the solid lines represent the JFP model predictions.}
\label{Figure_5}
\end{figure*}

 \subsection{Predictions from JFP Model}

Figs.~\ref{Figure_5}(a)-\ref{Figure_5}(f) shows the variation of the velocity $v_x$, shear rate $\dot{\gamma}$, solids fraction $\phi$, inertial number $I$, shear stress $\tau_{yx}$ and viscosity $\eta$ with distance $y$ from the base at different times for $\theta=24^\circ$ and restitution coefficient $e_n=0.5$. Black circles, red squares, and blue lower triangles represent the average DEM flow properties at $t=100$, $t=200$, and $t=500$ time units. The solid lines represent the predictions at different times obtained by numerically solving the Eq.~\ref{x_momentum_detail} using $PDEPE$ solver and using the JFP model (Eq.~\ref{mu_i_JFP}) for rheological description. The JFP model parameters obtained by fitting Eq.~\ref{mu_i_JFP} to the simulation data are shown in Table \ref{parameters_JFP_Model}.

Figure~\ref{Figure_5}(a) shows the variation of velocity $v_x$  with distance $y$ from the base at three different time instants of $t=100$, $t=200$, and $t=500$ time units. The velocity profile shows a Bagnold dependence with negligible slip at the base and increases to a maximum value at the free surface. The velocity at the free surface keeps increasing with time and the slope of the velocity profile near the base also changes. The shear rate near the base is maximum and decreases with an increase in height from the base as can be seen in Fig.~\ref{Figure_5}(b). The shear rate also increases with time at any given distance from the base.
Figure~\ref{Figure_5}(c) shows the inertial number $I$ along the height of the flowing layer at different time instants. As expected, the inertial number also increases with time and shows minor variations in the layer for early times. At later times, it becomes nearly constant in most of the bulk layer with small oscillations.

Figure~\ref{Figure_5}(d) shows the variation of solids fraction $\phi$ along the flowing layer at different time instants. The solids fraction $\phi$ shows large fluctuation across the flowing layer due to averaging over only a few snapshots. In addition, the usage of nearly monodisperse particles in the 2D simulation may also lead to particle layering that can cause strong variations in solids fraction $\phi$ in different bins. With increasing time, a small decrease in solids fraction from $\phi\simeq0.8$ to $\phi\simeq0.75$ is observed.  
Figure~\ref{Figure_5}(e) shows a linear variation of the shear stress $\tau_{yx}$ with distance $y$ from the base due to the nearly constant bulk density of the layer. Figure~\ref{Figure_5}(f) shows the viscosity $\eta$ with distance $y$ from the base. The viscosity varies non-linearly from zero at the free surface to a maximum value at the base. With the increase in time, the viscosity decreases. The theoretical predictions obtained using the JFP rheological model are indeed able to capture the flow properties at different times for $\theta=24^\circ$.
\vspace{1.7cm}

\begin{table}
\begin{ruledtabular}
\caption{\label{parameters_JFP_Model}Model parameters for JFP model}
\begin{tabular}{lcr}
Expression & \hspace{12cm} $e=0.5$ \\ 
\hline
\\
& \hspace{12cm} $\mu_s^{'}=0.23$ \\
$\mu(I)=\mu_{s}^{'}+\frac{\mu_{m}^{'}-\mu_{s}^{'}}{1+I_{0}'/I}$ & \hspace{12cm} $\mu_m^{'}=0.69$ \\
            & \hspace{12cm} $I_0'=0.20$ \\
\end{tabular}
\end{ruledtabular}  
\end{table}

\begin{figure*}[hbtp]
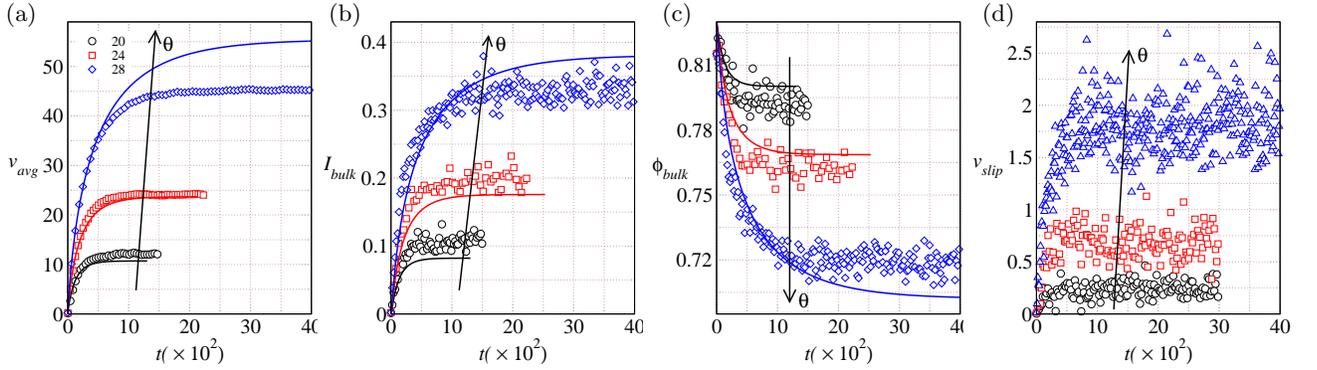

 	\includegraphics[scale=0.3]{Fig_6a.eps}
 	\put (-120,130) {(a)}
	\includegraphics[scale=0.3]{Fig_6b.eps}
	\put (-120,130){(b)}
	\includegraphics[scale=0.3]{Fig_6c.eps}
	\put (-117,130){(c)}
	\includegraphics[scale=0.3]{Fig_6d.eps}
	\put (-117,130){(d)}
\caption{Variation of the (a) average velocity $v_{avg}$, (b) average inertial number $I_{bulk}$, (c) average solids fraction in the bulk  $\phi_{bulk}$ and (d) slip velocity at the base $v_{slip}$ with time $t$ ($e_n=0.5$) for different inclinations $\theta$. Symbols represent the DEM simulations data whereas the solid lines in (a-c) represent the JFP model predictions.}
\label{Figure_6}
\end{figure*}

Figs.~\ref{Figure_6}(a)-\ref{Figure_6}(d) show the variation of the average velocity of the flowing layer $v_{avg}$, inertial number in the bulk $I_{bulk}$, average solids fraction in the bulk $\phi_{bulk}$ and the slip velocity at the base $v_{slip}$ with time $t$ for inclinations varying from low to moderate inclination angles. The average velocity $v_{avg}$ is calculated as $v_{avg}=\frac{1}{h}\int_{0}^{h}v_x dy$. 
As expected, the average velocity increases with time and eventually becomes constant at a steady state. The steady-state value of the average velocity increases with an increase in inclination.
Figs.~\ref{Figure_6}(b) and \ref{Figure_6}(c) show the variation of average inertial number and solids fraction in the bulk with time for different inclinations $\theta$. The average values in the bulk are calculated by considering data in the bulk region ($0.2h \leq y \leq 0.8h$) of the layer and discarding the data near the free surface and the base where $h$ is the free surface height.
The inertial number also increases with time and eventually reaches a steady state value and increases with the inclination angle. The bulk solids fraction decreases with time and inclination angle and eventually attains a steady state at large times.
\vspace{1.7cm}

\begin{figure*}[hbtp]
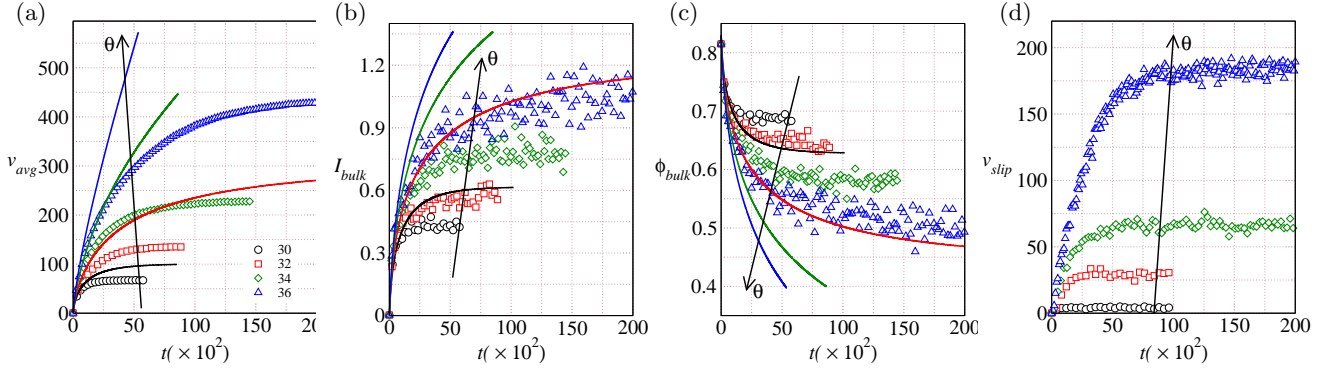

	\includegraphics[scale=0.3]{Fig_7a.eps}
 	\put (-120,130) {(a)}
	\includegraphics[scale=0.3]{Fig_7b.eps}
	\put (-119,130) {(b)}
	\includegraphics[scale=0.3]{Fig_7c.eps}
	\put (-118,130) {(c)}
	\includegraphics[scale=0.3]{Fig_7d.eps}
	\put (-118,130) {(d)}
		\caption{Variation of the (a) average velocity of the layer $v_{avg}$, (b) average inertial number in the bulk $I_{bulk}$, (c) average solids fraction in the bulk  $\phi_{bulk}$ and (d) slip velocity at the base $v_{slip}$ with time $t$ ($e_n=0.5$) for different inclinations. Symbols represent the DEM simulation data and the solid lines represent the JFP model predictions.}
\label{Figure_7}
\end{figure*}

Figure~\ref{Figure_6}(d) shows the variation of slip velocity $v_{slip}$ with time $t$ for low to moderate inclinations. The slip velocity increases with an increase in inclination over time and eventually attains a steady state with small fluctuations around the steady mean slip velocity. The slip velocity in the case of these inclination angles is very small and neglecting the slip velocity in the calculation of average velocity barely affects the results. The effect, however, becomes important at higher inclinations and hence almost all the results reporting instantaneous/average velocity account for the slip velocity at the base in this study. Only the results shown in Fig.~\ref{Figure_3}(b) assume the slip velocity at the base to be zero to remain consistent with the theoretical predictions of \citet{parez2016unsteady}. 

The solid lines in Fig.~\ref{Figure_6} represent the theoretical predictions obtained from the $PDEPE$ solver using the JFP model. Figs.~\ref{Figure_6}(a)-\ref{Figure_6}(c) show that the JFP model predicts the flow behavior very well for $\theta\leq24^\circ$. However, the JFP model predictions appear to be deviating from the DEM data for $\theta=28^\circ$. This can be attributed to the presence of normal stress difference which remains negligible for low inclinations but starts to become important as the angle increases \cite{tripathi2011rheology,patro2021rheology}.
Next, we consider even higher inclination angles and compare the predictions of the continuum model using JFP model with DEM simulation results.

Figs.~\ref{Figure_7}(a)-\ref{Figure_7}(d) show the variation of average velocity $v_{avg}$, bulk inertial number $I_{bulk}$, bulk solids fraction $\phi_{bulk}$ for inclinations varying from $\theta=30^\circ$ to $\theta=36^\circ$. As before, the DEM simulations show that the average velocity and bulk inertial number increase with time and inclination angle and eventually attain a constant value after a sufficiently long time indicating the existence of a steady state at all four inclination angles. 
\begin{figure*}[hbtp]
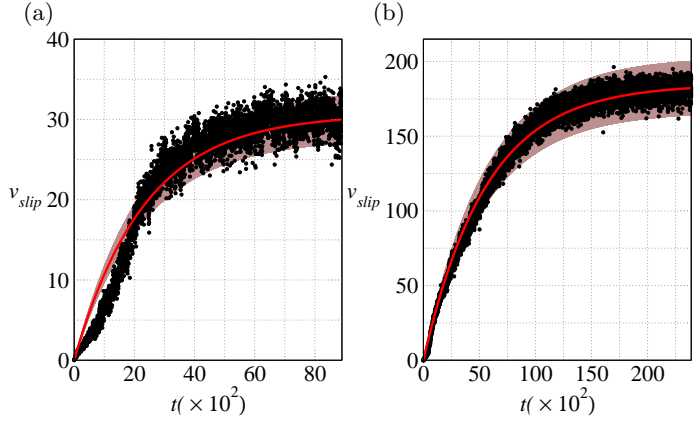

 	\includegraphics[scale=0.33]{Fig_8a.eps}
 	\put (-122,150) {(a)}
	\includegraphics[scale=0.33]{Fig_8b.eps}
	\put (-122,150){(b)}
	\caption{Variation of the slip velocity $V_{slip}$ with time $t$ for $e_n=0.5$ corresponding to (a) $\theta=32^\circ$ and (b) $\theta=36^\circ$. Black symbols represent the DEM simulation data. Red solid line represents the exponential fit to the DEM data. The Brown band represents the modeled variation of the slip velocity used in the theoretical predictions by accounting for the fluctuations in the mean slip velocity.}
\label{Figure_8}
\end{figure*}

 The predictions using the JFP model (solid lines) capture the flow properties for moderate angles very well but fail to capture the flow behavior for high inclinations. Specifically, the theoretical predictions using the JFP model, show continuously increasing average velocity and bulk inertial number for all $\theta \geq 32^\circ$ and do not achieve a steady state during the time period of interest. The difference between the predicted values and DEM results keeps increasing with the inclination angle and the theoretical predictions for the higher angles (shown in Fig.~\ref{Figure_7}) differ from the DEM results by a factor of two. 
Similar differences are observed in the predictions of the average inertial number $I_{bulk}$ and average solids fraction $\phi_{bulk}$ using the JFP model at higher inclinations. The predicted values of the bulk inertial number are found to be much higher and the bulk solids fractions are found to be substantially smaller than those observed in DEM simulations. 
Figure~\ref{Figure_7}(d) shows that the slip velocity in case of such high inclination angles is very significant. Neglecting such large slip velocity in the theoretical predictions affects the predicted values of the average velocity significantly.  

Figure \ref{Figure_8} shows the variation of the slip velocity $v_{slip}$ with time $t$ for two different angles $\theta=32^\circ$ and $\theta=36^\circ$. Although a small deviation from the fitted curve is evident at early times, the variation of the slip velocity with time is captured reasonably well using an exponential fit, shown using red solid lines in Fig.~\ref{Figure_8}. A large scatter in the instantaneous slip velocity values is evident around the mean exponential variation. This scatter appears to be within $5\%-10\%$ of the mean value obtained from the exponential fit. In the results shown in this study (except Fig.~\ref{Figure_3}), the theoretical predictions utilize the fitted exponential time variation of the slip velocity  while calculating the average velocity and instantaneous velocity profiles. The results for the shear rate, inertial number, packing fraction, shear stress, pressure, and the apparent viscosity are not affected due to the presence of the slip velocity. 
\vspace{1.7cm}

\begin{figure*}[hbtp]
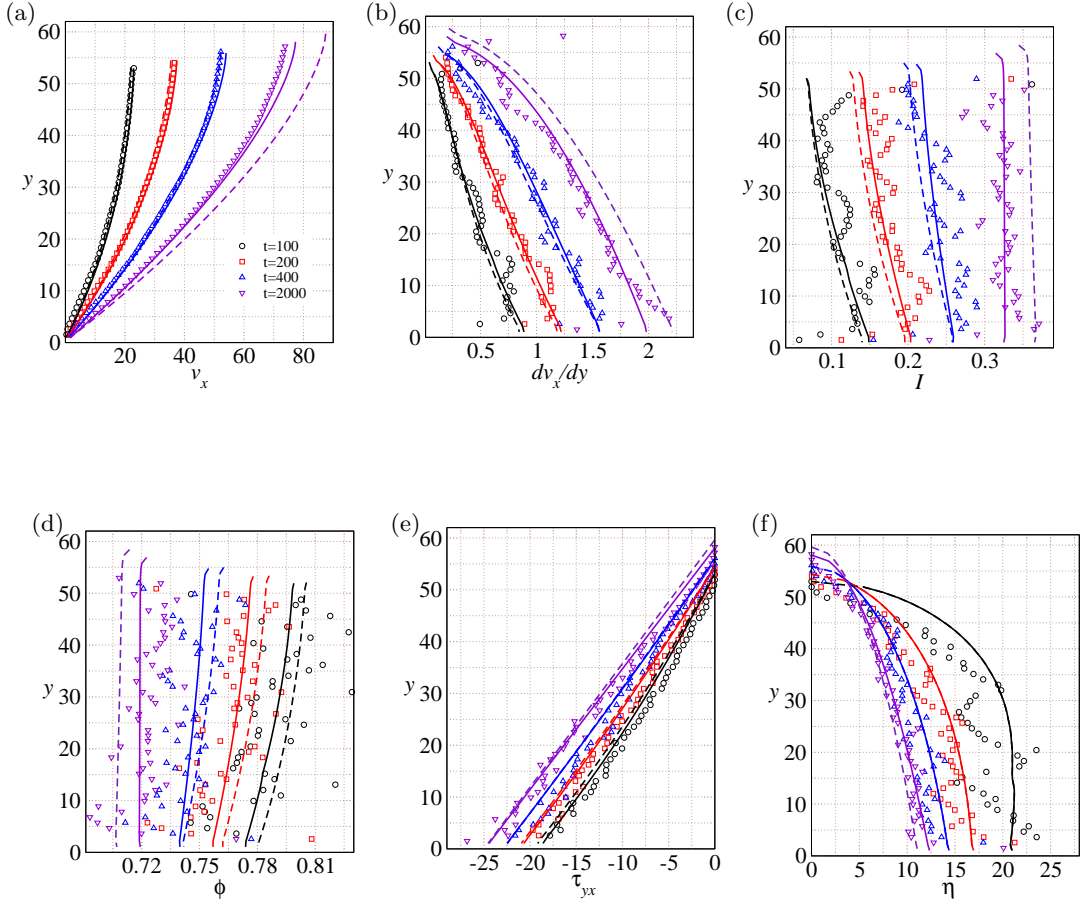

 	\includegraphics[scale=0.33]{Fig_9a.eps}
 	\put (-125,140) {(a)}
 	\hspace{0.5cm}
	\includegraphics[scale=0.33]{Fig_9b.eps}
	\put (-125,140) {(b)}
	\hspace{0.5cm}
	\includegraphics[scale=0.33]{Fig_9c.eps}
	\put (-125,140) {(c)}
    
	\vspace{1.6cm}
	\hspace{0.5cm}
	\includegraphics[scale=0.33]{Fig_9d.eps}
 	\put (-123,140) {(d)}
 	\hspace{0.5cm}
	\includegraphics[scale=0.33]{Fig_9e.eps}
	\put (-125,140) {(e)}
	\hspace{0.5cm}
	\includegraphics[scale=0.33]{Fig_9f.eps}
	\put (-125,140) {(f)}
	\caption{Variation of the (a) velocity $v_x$, (b) shear rate $\dot{\gamma}$, (c) inertial number $I$, (d) solids fraction $\phi$, (e) shear stress $\tau_{yx}$ and (f) viscosity $\eta$ with distance $y$ from the base at different times for $\theta=28^\circ$. Symbols represent the DEM simulations data ($e_n=0.5$), solid lines are the predictions of the modified rheology, and dashed lines represent the JFP model predictions.}
\label{Figure_9}
\end{figure*}

The results shown in this section confirm that the substantial difference observed between the DEM simulations and the theoretical predictions arise due to the inability of the JFP model to capture the rheology appropriately at high inclinations. In the next section, we show that the modified rheological description is able to capture the time-dependent properties in very good agreement with the DEM simulation results. 

\subsection{Predictions using modified rheological description}

\begin{figure*}[hbtp]
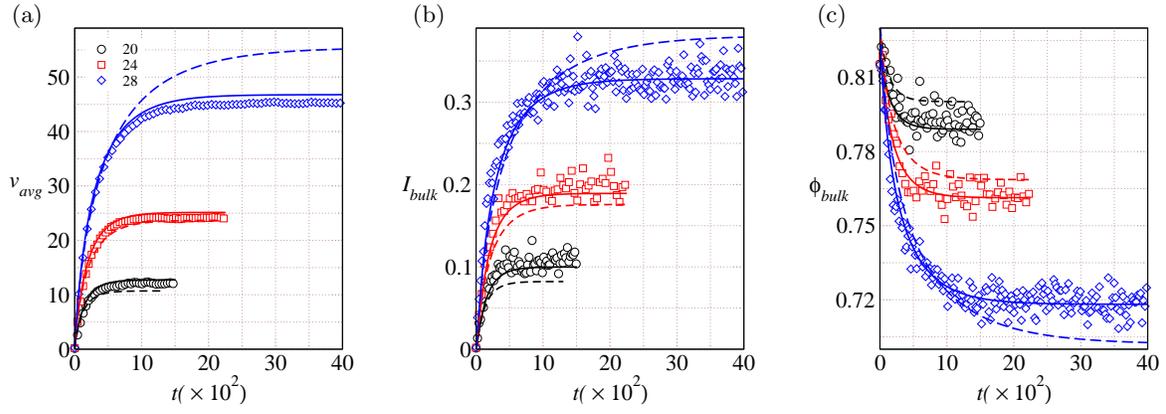

 	\includegraphics[scale=0.33]{Fig_10a.eps}
 	\put (-130,145) {(a)}
 	\hspace{0.5cm}
	\includegraphics[scale=0.33]{Fig_10b.eps}
	\put (-130,145){(b)}
	\hspace{0.5cm}
	\includegraphics[scale=0.33]{Fig_10c.eps}
	\put (-130,145){(c)}
	\caption{Variation of the (a) average velocity $v_{avg}$, (b) average inertial number $I_{bulk}$, and (c) average solids fraction $\phi_{bulk}$ with time $t$ for different inclinations $\theta$. Symbols represent the DEM simulations data ($e_n=0.5$). Symbols represent the DEM simulations data ($e_n=0.5$), solid lines are the predictions of the modified rheology and dashed lines represent the JFP model predictions.}
\label{Figure_10}
\end{figure*}
In this section, we report the predictions of the continuum model utilizing the new rheological description proposed by \citet{patro2021rheology} which compliments the MK model along with a normal stress difference law.
Figs.~\ref{Figure_9}(a)-\ref{Figure_9}(f) report the results for $\theta=28^\circ$ at four different time instances. Symbols represent the variation of  velocity $v_x$, shear rate $dv_x/dy$, inertial number $I$, solid fraction $\phi$, shear stress $\tau_{yx}$ and viscosity $\mu$ with distance $y$ from the base. Black circles represent the flow properties at $t=100$. Red squares represent the flow properties at $t=200$. Similarly, blue upper triangles and purple lower triangles represent the flow properties at $t=400$ and $t=2000$ respectively. The solid lines represent the theoretical predictions obtained from the continuum model using the modified rheology of \citet{patro2021rheology}. 
The rheological model parameters for the modified $\mu-I$ relation, dilatancy law and normal stress difference law are shown in Table \ref{parameter_mk_model}, Table \ref{parameters_dilatancy_power_law} and Table \ref{parameter_n1_by_p_quad}, respectively. The predictions using the modified rheology are able to capture the flow properties very well at different times for $\theta=28^\circ$. For comparison, we also show the predictions obtained using the JFP model using dashed lines. The predictions for the JFP model differ marginally from the modified rheology predictions and DEM data for early times ($t\leq1000$). At larger times, however, the JFP model predictions differ substantially from the DEM data.

\begin{table}[htbp]
\caption{\label{parameter_mk_model}Parameters for MK model}
\begin{ruledtabular}
\begin{tabular}{ccccc}
$e$ & $\mu_s$	& $c_1$ & $c_2$	& $I_0$\\  
\hline
0.1  & 0.26 & 0.92 & 0.36 & 0.69\\ 0.5  & 0.24 & 0.68 & 0.22 & 0.39\\ 
\end{tabular}
\end{ruledtabular}
\end{table}

\begin{table}[htbp]
\caption{\label{parameters_dilatancy_power_law}Model parameters for dilatancy law}
\begin{ruledtabular}
\begin{tabular}{cccc}
$e$ & $\phi_{max}$ & $a$ & $\alpha$ \\ \hline
0.1 & 0.82 & 0.30 & 0.99 \\
0.5 & 0.82 & 0.31 & 1.00 \\
\end{tabular}
\end{ruledtabular}
\end{table}

\begin{table}[htbp]
\caption{\label{parameter_n1_by_p_quad}Model parameters for normal stress difference law}
\begin{ruledtabular}  
\begin{tabular}{lllcc}
\hspace{3.5cm}$I>I^*=0.1$\hspace{10cm}$I\leq I^*=0.1$\\
\hline
$e$\hspace{2.1cm}$A$\hspace{1.5cm}$B$ \hspace{1.5cm}$C$\hspace{10.1cm}$K$ \\ \hline
0.1\hspace{1.7cm}0.16\hspace{1.2cm}0.39\hspace{1.2cm}-0.1\hspace{9.7cm}-0.06 \\

0.5\hspace{1.7cm}0.10\hspace{1.2cm}0.43\hspace{1.2cm}-0.1\hspace{9.7cm}-0.06 \\

\end{tabular}
\end{ruledtabular}
\end{table}

\begin{figure*}[hbtp]
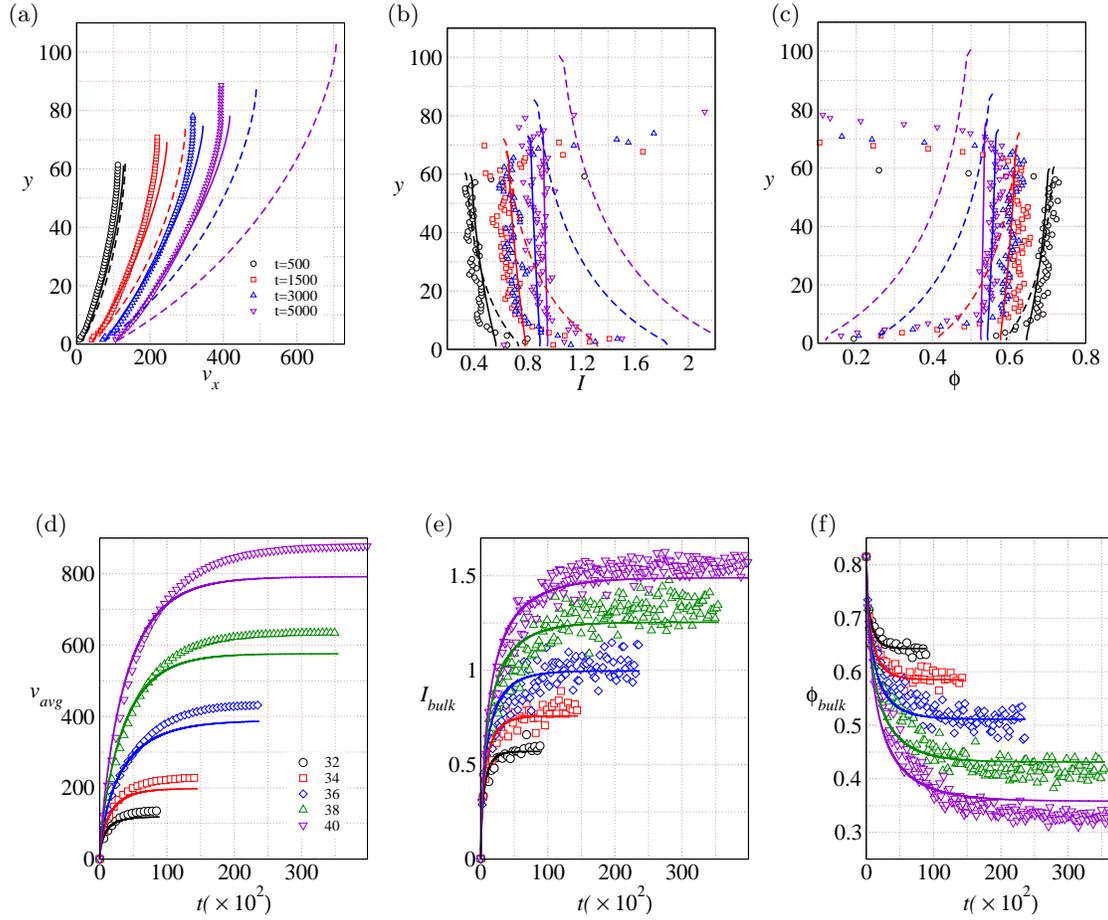

	\includegraphics[scale=0.33]{Fig_11a.eps}
 	\put (-128,140) {(a)}
 	\hspace{0.5cm}
	\includegraphics[scale=0.33]{Fig_11b.eps}
	\put (-125,140){(b)}
	\hspace{0.5cm}
	\includegraphics[scale=0.33]{Fig_11c.eps}
	\put (-125,140){(c)}
    
	\vspace{1.6cm}
    \hspace{0.5cm}
    \includegraphics[scale=0.33]{Fig_11d.eps}
	\put (-128,145){(d)}
    \hspace{0.5cm}
	\includegraphics[scale=0.33]{Fig_11e.eps}
	\put (-125,145){(e)}
    \hspace{0.5cm}
	\includegraphics[scale=0.33]{Fig_11f.eps}
    \put (-125,145){(f)}
	\caption{Variation of the (a) velocity $v_x$, (b) inertial number $I$, (c) solids fraction $\phi$ with distance $y$ from the base at different times for $\theta=36^\circ$. Variation of the (d) average velocity $v_{avg}$, (e) average inertial number $I_{bulk}$ and (f) average solids fraction $\phi_{bulk}$ with time $t$ for even inclinations $32^\circ\leq\theta\leq40^\circ$. Symbols represent the DEM simulations data ($e_n=0.5$), solid lines are the predictions of the modified rheology and dashed lines represent the JFP model predictions.}
\label{Figure_11}
\end{figure*}

This deviation of JFP model predictions at larger times becomes more evident from Figs.~\ref{Figure_10}(a)-\ref{Figure_10}(c). These figures show the variation of average velocity $v_{avg}$, bulk inertial number $I_{bulk}$, and bulk solids fraction $\phi_{bulk}$ with time $t$ for three different inclinations. In all the cases, the predictions from the modified rheology (shown as solid lines) agree with the DEM simulations data better compared to the JFP model predictions shown using the dashed lines. While predicting the average velocity in Fig.~\ref{Figure_10}(a), we account for the slip velocity at the base as mentioned before. However, assuming the slip velocity to be zero leads to very small differences in the flow predictions at these angles. 

Figs.~\ref{Figure_11}(a)-\ref{Figure_11}(c) show the variation of velocity $v_x$, inertial number $I$, and solids fraction $\phi$ with the distance from the base $y$ using the modified rheology for $e_n=0.5$ at $\theta=36^\circ$. The flow profiles observed in DEM simulations are indeed captured very well using the modified rheology (shown using solid lines). The JFP model, on the other hand, has serious limitations in predicting the flow properties (shown using dashed lines) accurately at such high inclinations. Specifically, the JFP model predictions lead to significant over-predictions of the velocity and inertial number and under-prediction of the solids fraction at large times. Figs.~\ref{Figure_11}(d)-\ref{Figure_11}(f) show the variation of average velocity $v_{avg}$, average inertial number $I_{bulk}$ and average solids fraction $\phi_{bulk}$ with time $t$ for stiff inclinations in the range $32^\circ\leq\theta\leq40^\circ$. The solid lines represent the predictions of the modified rheology \citep{patro2021rheology}. For the higher values of inclination angles, the average inertial number exceeds unity. However, the modified rheological model is able to predict the transient flow properties in reasonable agreement with the DEM simulations.

\begin{figure*}
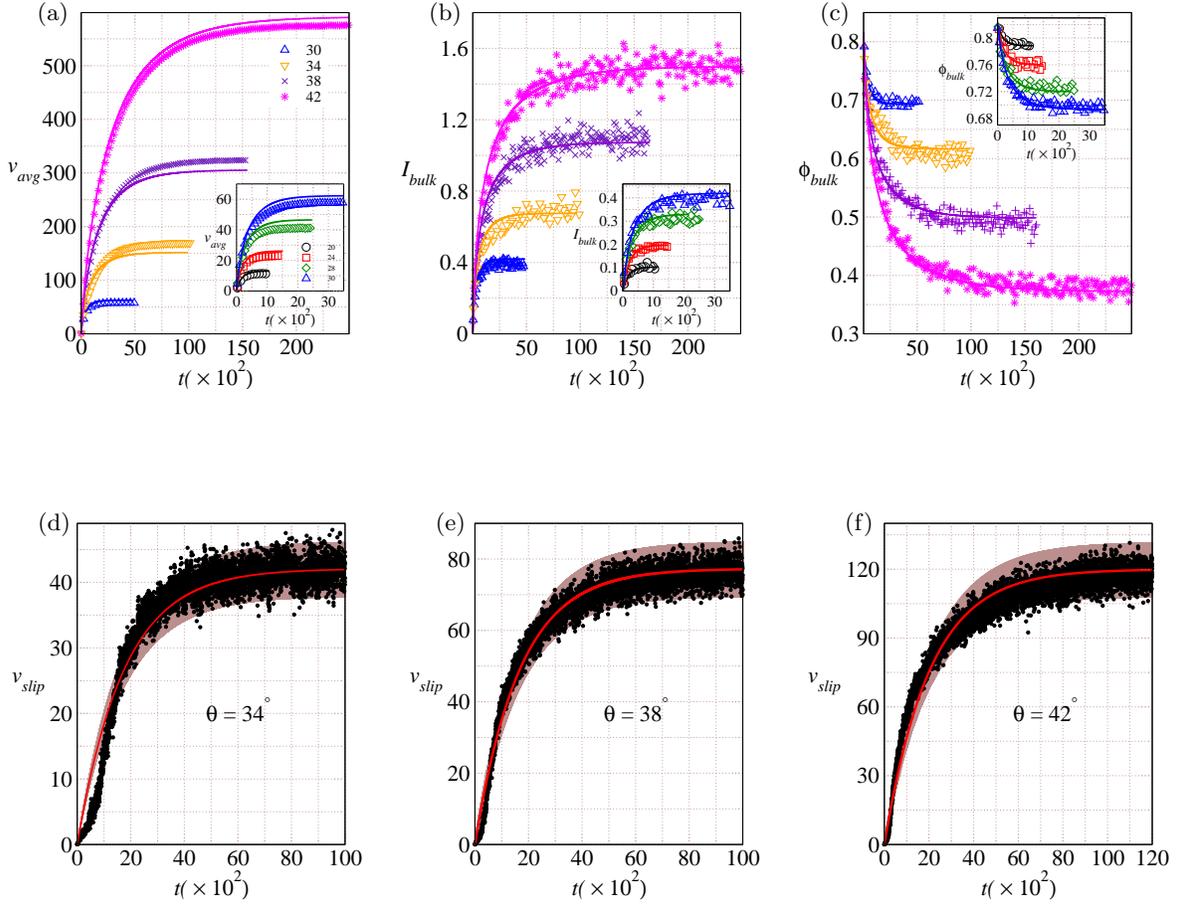

 	\includegraphics[scale=0.33]{Fig_12a.eps}
 	\put (-120,140) {(a)}
 	\hspace{0.5cm}
	\includegraphics[scale=0.33]{Fig_12b.eps}
	\put (-120,140){(b)}
	\hspace{0.5cm}
	\includegraphics[scale=0.33]{Fig_12c.eps}
	\put (-120,140){(c)}
	\hspace{0.5cm}
    
	\vspace{1.6cm}
	\includegraphics[scale=0.33]{Fig_12d.eps}
 	\put (-123,140) {(d)}
 	\hspace{0.5cm}
	\includegraphics[scale=0.33]{Fig_12e.eps}
	\put (-123,140){(e)}
	\hspace{0.5cm}
	\includegraphics[scale=0.33]{Fig_12f.eps}
	\put (-123,140){(f)}
	\caption{Variation of (a) the average velocity $v_{avg}$, (b) average inertial number $I_{bulk}$, (c) average solids fraction $\phi_{bulk}$ with time $t$ at different inclinations $\theta$. Symbols represent the DEM data for $e_n=0.1$ whereas the solid lines represent the theoretical predictions obtained from the modified rheology. Inset shows the results for low to moderate inclinations. The results for higher inclination angles are reported in the main figure. Slip velocity at the base $v_{slip}$ as a function of time for inclination (d) $\theta=34^\circ$, (e) $\theta=38^\circ$ and (f) $\theta=42^\circ$. Red line shows the exponential fit to the simulation data (shown using black circles). Brown band shows the variation of $5\%$ from the mean velocity and is able to capture the slip velocity data at all time.}
\label{Figure_12}
\end{figure*}

Figs.~\ref{Figure_12}(a)-\ref{Figure_12}(c) report the results for highly dissipative particles with $e_n=0.1$ spanning a large range of inclinations. Symbols represent the DEM data and the solid lines represent the modified rheological model predictions by \citep{patro2021rheology}. Figs.~\ref{Figure_12}(a)-\ref{Figure_12}(c) report the average flow properties for moderate to high inclinations (i.e. $\theta=30^\circ,34^\circ,38^\circ$ and $42^\circ$). The results for inclinations $\theta\leq30^\circ$ are shown in the inset. The steady state average velocity varies from $v_{avg}^{ss}$ $\sim$ 10 at $\theta=20^\circ$ to $v_{avg}^{ss}$ $\sim$ 580 at $\theta=42^\circ$. The inertial number varies from around $I_{avg}$ $\sim$ 0.1 at lowest inclination to $I_{avg}$ $\sim$ 1.5 for highest inclination. The modified rheology is able to predict the entire range of velocity profiles, inertial number and solids fraction observed across these different tilt angles accurately. As mentioned before, accurate prediction of velocity at high inclination requires reliable knowledge of the slip velocity at the base. These slip velocities for the three higher angles $\theta=34^\circ$, $\theta=38^\circ$ and $\theta=42^\circ$ are reported in Figs.~\ref{Figure_12}(d),\ref{Figure_12}(e) and \ref{Figure_12}(f) respectively. Despite small deviations from the mean trend, an exponential variation describes the slip velocity dependence on time very well. All the data seem to be within $5\%$ of the fitted mean slip velocity. In the predictions of properties shown in Figs.~\ref{Figure_12}(a)-\ref{Figure_12}(c), the fitted mean slip velocity with time is accounted for. This accounting of the mean slip velocity is crucial for accurate predictions of the average and instantaneous velocities. Other properties, however, do not get altered by ignoring the slip velocity at the base.

\section{\label{sec:Discussion}Discussion}

\begin{figure*}[hbtp]
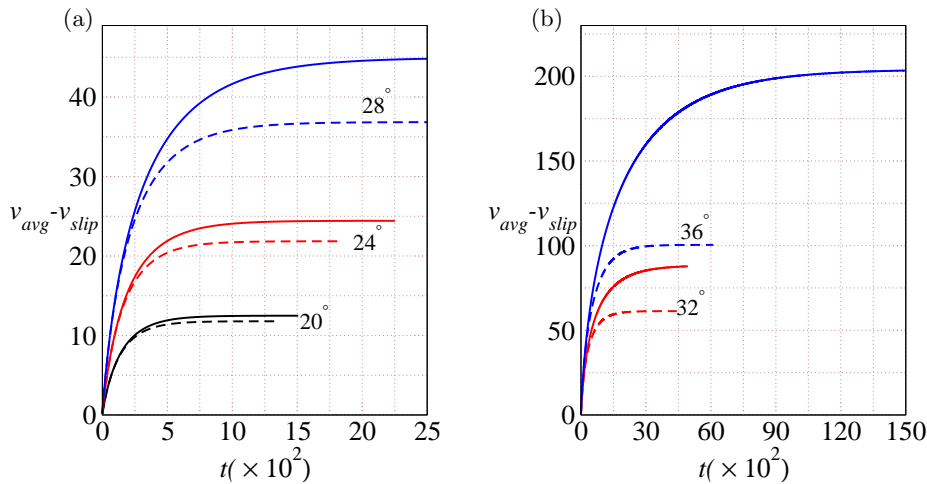

 	\includegraphics[scale=0.4]{Fig_13a.eps}
 	\put (-143,173) {(a)}
 	\hspace{0.5cm}
	\includegraphics[scale=0.4]{Fig_13b.eps}
	\put (-150,173) {(b)}
	\hspace{0.5cm}
	\caption{Variation of the average velocity $v_{avg}$ with time $t$ for $e_n=0.5$ at (a) 
	$\theta=20^\circ$, $24^\circ$ and $28^\circ$ (b) $\theta=32^\circ$ and $36^\circ$. The lines are the theoretical predictions from the modified rheological model. Solid lines are predictions considering variable flowing layer thickness, while dashed lines are predictions considering a constant flowing layer thickness.}
	\label{Figure_13}
\end{figure*}

\begin{figure*}[hbtp]
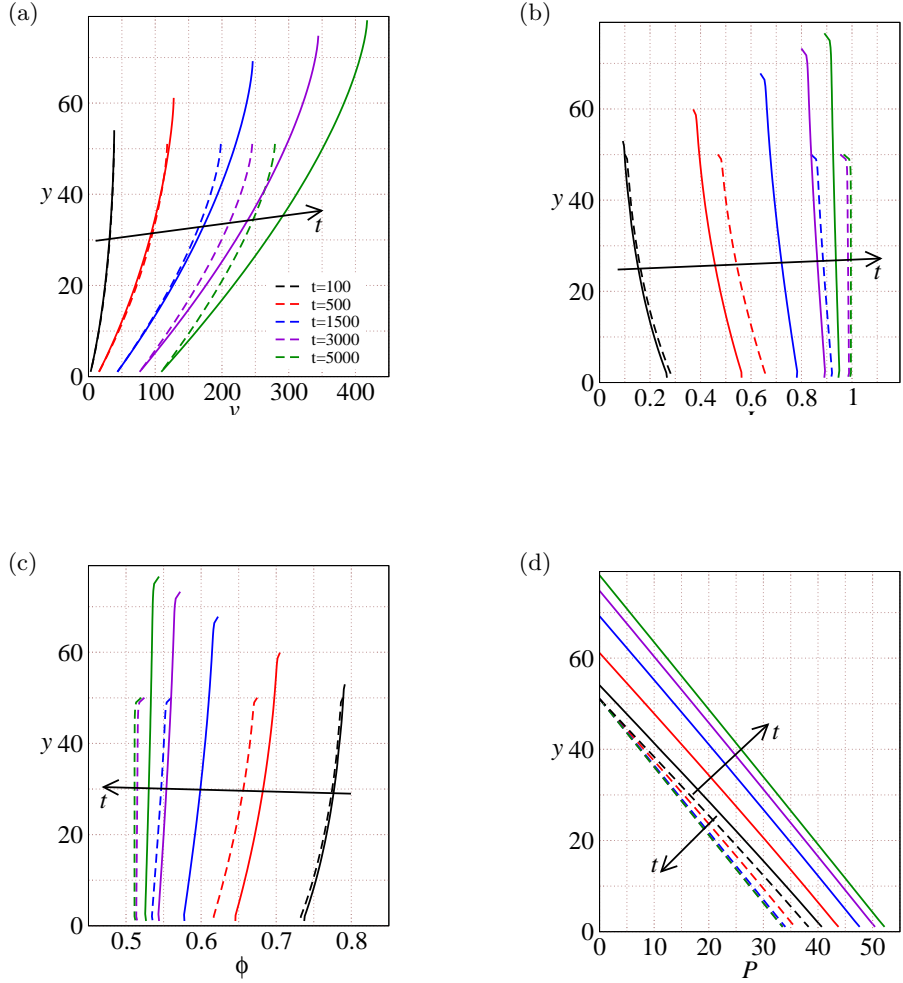

 \includegraphics[scale=0.37]{Fig_14a.eps}
    \put (-145,155){(a)}
    \hspace{2cm}
    \includegraphics[scale=0.37]{Fig_14b.eps}
	\put (-145,155){(b)}
    
	\vspace{1.6cm}
	\includegraphics[scale=0.37]{Fig_14c.eps}
    \put (-145,155){(c)}
	\hspace{2cm}
	\includegraphics[scale=0.37]{Fig_14d.eps}
	\put (-145,155){(d)}
	\caption{Variation of the (a) velocity $v_x$, (b) inertial number $I$, (c) solids fraction and (d) Pressure $P$ with distance $y$ from the base at different times for $e_n=0.5$ at $\theta=36^\circ$. The solid lines are the theoretical predictions from the rheological model considering variable flowing layer thickness. The dashed lines are the theoretical predictions from the rheological model considering constant flowing layer thickness.}
\label{Figure_14}
\end{figure*}

\subsection{Importance of granular dilatancy}
The modified $\mu-I$ rheological model along with a normal stress difference law is able to predict the flow profiles at large inclination angles that are in good agreement with the DEM simulation results.
However, this accurate prediction requires the continuum approach to keep track of the free surface. In this work, we have utilized a simplified method to account for the variation of the flowing layer thickness $h(t)$ with time $t$. Using the mass balance and equating the total mass at any instant to be equal to the mass at $t=0$, the flowing layer thickness $h$ at every instant is updated by $h(t)=h_{min}\phi_{max}/\phi_{avg}(t)$. Most of the studies dealing with granular rheology assume the flow to be incompressible and hence do not bother about such dilation effects. In order to check if this increase of the layer height with time is critical to get accurate flow properties, we also perform continuum simulations assuming that the flowing layer thickness remains constant for the entire range of inclination angles. Note that, the continuum predictions for the velocity obtained by solving Eq.~\ref{x_momentum_detail} do not predict any slip velocity (which is added to the predictions for comparison with DEM data), we compare the predictions for $v_{avg}-v_{slip}$ with time $t$ in Fig.~\ref{Figure_13}.

Figure~\ref{Figure_13}(a) shows the variation of $v_{avg}-v_{slip}$ for $\theta=20^\circ$, $24^\circ $ and $28^\circ$. The average velocity predicted by assuming a constant flowing layer thickness is close to the predictions for variable flowing layer thickness at low inclinations. As the angle of inclination increases, the deviation between the constant flowing layer thickness case and the variable flowing layer thickness case starts to increase as evident in Fig.~\ref{Figure_13}(b).
These results confirm that the effect of the layer dilation becomes crucial and needs to be accounted at high inclinations. This can be seen more clearly in Figs.~\ref{Figure_14}(a)-\ref{Figure_14}(d) where the constant layer thickness predictions start deviating from the variable layer thickness predictions at $t=500$ and become much larger at higher times. Note that the velocity (shown in Fig.~\ref{Figure_14}(a)) across the layer for the two cases at $t=500$ differs only near the free surface. However, the inertial number and the solids fraction at $t=500$ for the two cases differ significantly all across the layer. This noticeable difference between the two cases can be understood in the following manner. With the increase in the velocity, the shear rate, and hence, the inertial number $I$ increases with time. This increase in $I$ causes the solids fraction $\phi$ (and hence bulk density $\rho_b=\phi \rho_p$) to decrease. 
In the constant height case, the reduction in the bulk density leads to a small reduction in the pressure (dashed lines in Fig.~\ref{Figure_14}(d)) with time. Since $I=\dot{\gamma}d/(\sqrt{P/\rho_p})$, this small reduction in pressure contributes to the increase of the inertial number only marginally, and most of the contribution to the inertial number increase is due to the change in the shear rate.
However, when the dilation of the layer is allowed, the height of the layer increases and hence the pressure at all locations in the layer increases with time (solid lines in Fig.~\ref{Figure_14}(d)). This increased pressure leads to a reduction in the  inertial number while the shear rate increase leads to an increase in the inertial number. 
Figure~\ref{Figure_14}(b) shows that the net effect of these two competing influences of shear rate and pressure effectively leads to an increase of $I$. However, this increment in $I$ is smaller compared to the constant height case due to the increase in pressure. With time, the inertial number increases and the change in pressure become even more prominent and the discrepancy between the variable and constant layer height case keeps increasing.

\begin{figure*}[hbtp]
 	\includegraphics[scale=0.44]{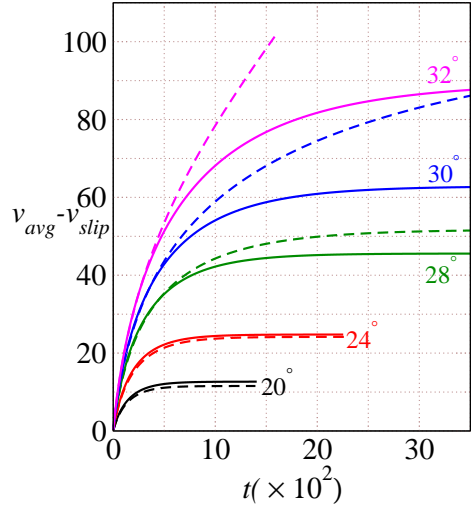}
		\caption{Variation of the average velocity $v_{avg}$ with time $t$ for $e_n=0.5$ at $\theta=20^\circ$, $24^\circ$, $28^\circ$, $30^\circ$ and $32^\circ$. Solid lines are predictions considering normal stress difference, while dashed lines are predictions without considering the effect of normal stress difference. }
\label{Figure_15}
\end{figure*}

\subsection{Role of normal stress difference}
One of the key features of the rheology proposed by \citet{patro2021rheology} is the significant role of the normal stress difference due to the anisotropy in the diagonal components of the stress tensor. This stress anisotropy, along with the non-monotonic-$\mu-I$ relation makes the rheological description complete. The JFP model, on the other hand, ignores the stress anisotropy and assumes the normal stress difference to be zero. In order to explore the importance of normal stress differences in predicting the flow properties accurately, we predicted the average velocity by ignoring the normal stress differences and using only the non-monotonic-$\mu-I$ variation. As shown in Fig.~\ref{Figure_13}, the slip velocity has not been included in these predictions. These predictions are found to be nearly identical to the predictions accounting for the presence of the normal stress difference for $\theta\leq24^\circ$. This is due to the fact that the normal stress difference to pressure ratio is very small at low inclinations (i.e. at low inertial numbers). However, at higher inclinations, the difference between the solid and dashed curves of the same color becomes significant (Fig.~\ref{Figure_15}) confirming that the presence of normal stress difference cannot be ignored at higher inclinations i.e., at $\theta=30^\circ$ and $\theta=32^\circ$. These results indicate that accounting for the non-monotonic variation of $\mu-I$ alone is not enough to capture the flow properties accurately for flows at high inertial numbers (observed at high inclinations).  
\vspace{1.7cm}

\begin{figure*}[hbtp]
\includegraphics[scale=0.33]{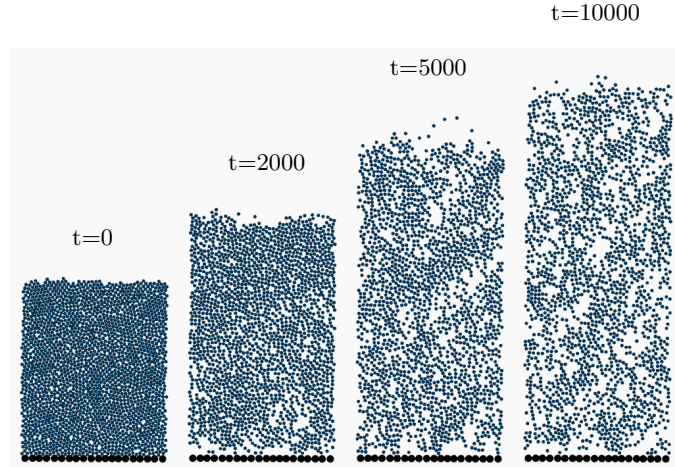}
\put(-230,85){t=0}
\put(-171,113){t=2000}
\put(-110,149){t=5000}
\put(-49,170){t=10000}
\caption{Snapshots of system's configuration at different times for $e_n=0.5$ at $\theta=42^\circ$.}
\label{Figure_16}
\end{figure*}

\subsection{Oscillations in the steady flow}

Figure~\ref{Figure_16} shows the snapshots of the flowing  layer ($e_n=0.5$) for an inclination of $\theta=42^\circ$ at four different time instants. Starting from a packed arrangement of particles at $t=0$, the layer height increases as the flow evolves with time and achieves more than twice of the initial layer height. The reduction in the solids fraction accompanied with this layer height increase is evident from the sparse arrangement of particles in the bed. Note that the region near the base depicts lower solids fraction compared to the bulk of the layer. In addition, the arrangement of particles in the layer shows regions of varying density along the flow as well as normal direction. These density variations at high angles are observed along with noticeable fluctuations in the layer height. 
\vspace{1.7cm}

\begin{figure*}[hbtp]
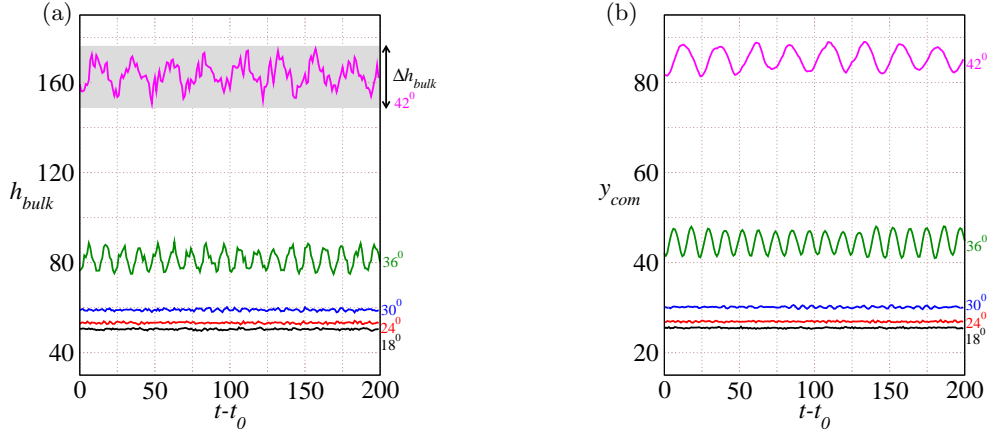

 	\includegraphics[scale=0.37]{Fig_17a.eps}
 	\put (-150,155) {(a)}
 	\hspace{2cm}
	\includegraphics[scale=0.37]{Fig_17b.eps}
	\put (-145,155) {(b)}
	\caption{Steady state oscillating behavior of a) bulk flowing layer thickness $h_{bulk}$ 
    and b) center of mass $y_{com}$ at different inclinations for restitution coefficient $e_n=0.5$.}
\label{Figure_17}
\end{figure*}

These oscillations in the bulk layer height $h_{bulk}$ are shown in Fig.~\ref{Figure_17}(a) over a time period of $200$ time units starting from a time instant $t_0$. 
Oscillations in the center of mass $y_{com}$ are shown in Fig.~\ref{Figure_17}(b). 
These height measurements are done after the flow has achieved steady kinetic energy (hence $t_0$ varies with the inclination angle $\theta$). 
 The oscillations in the bulk layer height $h_{bulk}$ as well as center of mass $y_{com}$ keep increasing with the inclination angle. While the difference between maximum and minimum bulk height is around a couple of particle diameters for inclinations $\theta\leq 30^\circ$, this variation becomes as large as $\Delta h_{bulk}\approx 16d$ for $\theta= 42^\circ$ case (Fig.~\ref{Figure_17}(a)). As shown in Fig.~\ref{Figure_17}(b), the amplitude of the oscillations observed in the center of mass position is approximately half of that observed in the bulk layer height. These large amplitude oscillations in the layer height at high inclinations might be linked to the pressure (or expansion) waves that cause density variations in the system. Such variations are observable in the video link given in the supplementary information (see SI 1). These effects indicate that the role of density and height variations become important and cannot be ignored for granular flows at inertial numbers comparable to unity. Such oscillations at high chute inclinations have been mentioned by \citet{brodu_delannay_valance_richard_2015} as well. 
Due to these oscillations, the calculation of rheological parameters in \cite{patro2021rheology} from simulation data at high inclinations requires averaging over a large number of snapshots so that the influence of these fluctuations is averaged out. 
Further, the data near the free surface and chute base are ignored during the rheological parameter estimation as \citet{mandal2016study} have shown that the $\mu-I$ data near the free surface and base follow slightly different behavior as compared to the data in the bulk region. 

In classical fluids, as the flow becomes faster, the competition between the inertia and gravity leads to instability and the free surface shows long wave modulations. In the case of turbulent flows, this instability is known as roll wave instability while in the case of viscous fluids, it is referred to as Kapitza instability. Such instabilities have been experimentally observed in the case of granular media before \cite{forterre_instability_2002}. Since our simulations use a periodic box, such long wave undulations are not observable. However, the oscillations in layer height (as shown in Fig.~\ref{Figure_17}(a)) might be indicative of such instability in the flow. A pronounced rippling behavior in the case of gravity-driven free surface flow of Newtonian fluids occurs in the range of Reynolds numbers $20 \leq Re \leq 1500$ \cite{BSL_2007}. Beyond this Reynolds number, the flow of Newtonian fluids becomes chaotic and is considered to be turbulent. 

\begin{figure*}[hbtp]
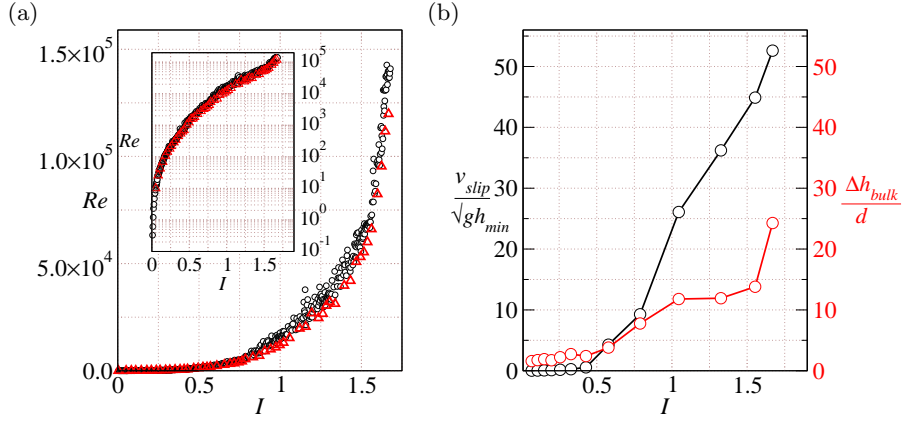

 	\includegraphics[scale=0.35]{Fig_18a.eps}
 	\put (-150,150) {(a)}
 	\hspace{0.5cm}
	\includegraphics[scale=0.35]{Fig_18b.eps}
	\put (-180,150) {(b)}
	\hspace{0.5cm}
	\caption{a) Variation of the Reynolds number with the inertial number for $e_n=0.5$. The inset figure in (a) shows the variation of Reynolds number ($Re$) with the inertial number ($I$) in the logarithmic scale. Black circles correspond to $h_{max}$ whereas the red upper triangles correspond to $h_{bulk}$ for the layer thickness in Reynolds number calculation. b) Variation of the slip velocity $v_{slip}/\sqrt{gh_{min}}$ and $\Delta h_{bulk}/d$ (on the right ordinate) with the inertial number at steady state.}
	\label{Figure_18}
\end{figure*}

Since the granular flow behavior is controlled by inertial number, we plot the variation of the Reynolds number of the layer with the inertial number in Fig.~\ref{Figure_18}(a). Calculation of the Reynolds number for a fluid requires knowledge of its density and viscosity. While the bulk density of the flowing granular fluid remains constant across most of the layer, the viscosity keeps changing across the layer (see Fig.~\ref{Figure_9}(f)). In absence of a proper definition of Reynolds' number for granular flows, we use the average density $\rho_{avg}$ and average viscosity $\eta_{avg}$ of the layer by averaging these properties in the bulk region, i.e., the region away from the base and the free surface. The average velocity $v_{avg}$ is calculated by averaging across the entire flowing layer.
Using these average values, Reynolds number for the layer is calculated as $Re=\frac{\rho_{avg}v_{avg}h}{\eta_{avg}}$. 
The calculation of Reynolds number using $h=h_{max}$ is shown using black circles whereas the calculation based on $h=h_{bulk}$ is shown using red upper triangles in Fig.~\ref{Figure_18}(a). The difference between the two seems to become significant only at large inertial numbers. 

Figure~\ref{Figure_18}(b) shows the variation of scaled slip velocity $v_{slip}/(gh_{min})^{0.5}$ with the inertial number, where $h_{min}$ is the minimum layer thickness in the packed configuration at the beginning of the simulation. The scaled slip velocity becomes significantly different from zero for $I>0.5$. The difference in the maximum and minimum value of the bulk flowing layer thickness ($\Delta h_{bulk}/d)$, shown on the right ordinate) remains nearly constant for $I<0.5$ and shows a noticeable increase for higher inertial numbers. A close look at the inset of Fig.~\ref{Figure_18}(a), which shows the main graph on a log-linear scale, indicates that $I\approx 0.5$ corresponds to the $Re\approx1500$. As mentioned before, this limit of Reynolds number corresponds to the onset of turbulence in case of Newtonian fluids. Whether the noticeable changes in the slip velocity and the bulk layer height oscillations observed in the flow correspond to onset of turbulence in the flowing granular layer is an open question that remains to be explored. For the sake of completeness, it is worth mentioning that the limit $Re=20$ for the onset of rippling flow corresponds to $I\approx 0.07$. 
\vspace{5.1cm}

\begin{figure*}[h!]
 	\includegraphics[scale=0.43]{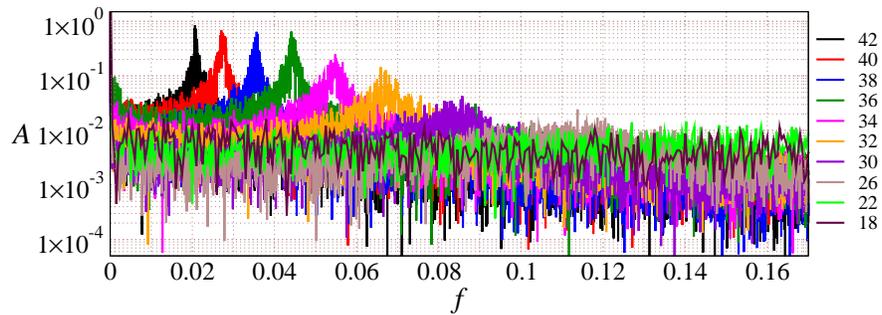}
\caption{Fast Fourier transform analysis of the center of mass data for all inclination angles. }
\label{Figure_19}
\end{figure*}

\begin{figure*}[hbtp]
 	\includegraphics[scale=0.36]{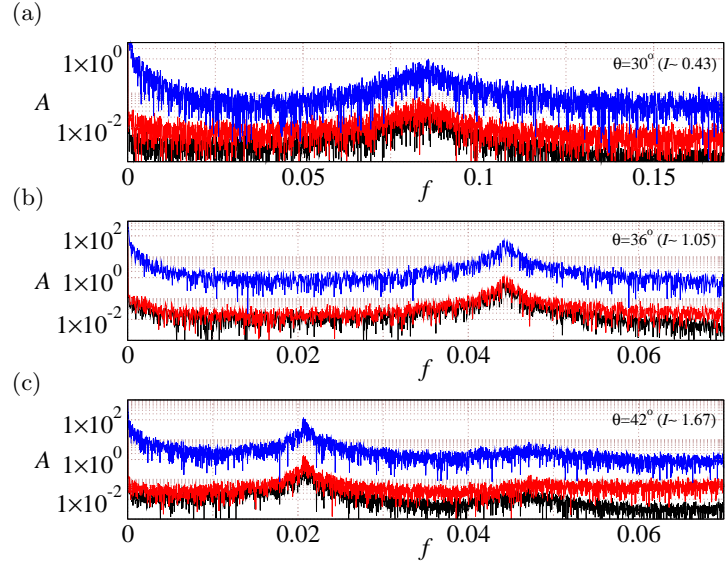}
 	\put (-270,205) {(a)}
 			\put (-270,135) {(b)}
 				\put (-270,65) {(c)}
    \caption{Fast Fourier transform analysis of the center of mass (shown in black line), bulk flowing layer thickness (shown in red line) and kinetic energy (shown in blue line) for inclination angle a) $\theta=30^\circ$, b) $\theta=36^\circ$ and c) $\theta=42^\circ$. }
\label{Figure_20}
\end{figure*}

In order to investigate the time periodic behavior of the layer at high inertial numbers, we performed a Fast Fourier Transform (FFT) analysis of the time series data of the center of mass position at steady state. 
Figure~\ref{Figure_19} shows the amplitude spectrum of the center of mass for different inclinations. 
The amplitude spectrum for inclinations $\theta=18^\circ$ and $\theta=22^\circ$ shows nearly uniform distribution for all frequencies. At higher inclinations, a dominant frequency with a clear peak starts to appear. 
The occurrence of a dominant frequency in the amplitude spectrum confirms that the variation of layer height occurs with a characteristic time period at these large inclinations. The peak frequency keeps moving to lower values as the inclination angle increases, indicating that the time period of oscillations increases with the inclination angle. Figure~\ref{Figure_19} also shows that the amplitude of oscillations at the $\theta=42^\circ$ is nearly two orders of magnitude higher compared to those at $\theta=26^\circ$. Hence the oscillations observed in the layer at high inclinations are of larger amplitude and occur over larger time periods compared to those at moderate inclinations. 
\vspace{1.7cm}

\begin{figure*}[hbtp]
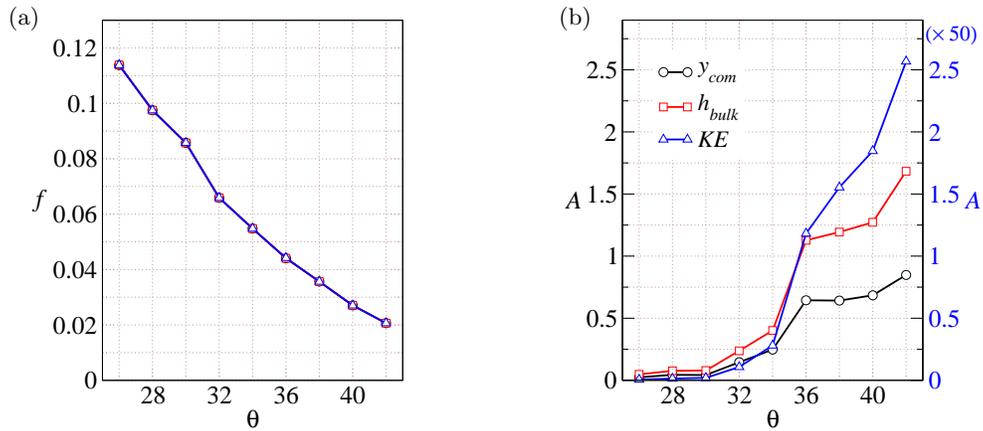

	\includegraphics[scale=0.37]{Fig_21a.eps}
    \put (-150,153){(a)}
		\hspace{2cm}
	\includegraphics[scale=0.37]{Fig_21b.eps}
		\put (-160,153){(b)}
		
	\caption{a) Variation of the dominant frequency $f$ corresponding to different inclination angles $\theta$. b) Variation of the amplitude $A$ corresponding to the dominant frequency $f$ (shown in (a)) for different inclination angles. See text for details} 
\label{Figure_21}
\end{figure*}

FFT analysis of the bulk layer height and the kinetic energy data are shown for three different inclinations in Fig.~\ref{Figure_20} using red and blue lines respectively. The black lines in the figure correspond to the center of mass position. The maximum amplitude of the spectrum is observed at the same frequency for all three properties. Hence we conclude that the oscillations at high inclinations affect these properties of the flow in a nearly identical fashion. This is more clearly observed in Fig.~\ref{Figure_21}(a) which shows that the dominant frequency for all three properties is identical at any given inclination. The figure also shows that the oscillation frequency decreases with an increase in the inclination angle. Figure~\ref{Figure_21}(b) shows that the amplitude of oscillations increases with the inclination angle and the amplitude of the oscillations in the bulk layer height is approximately twice of that in the center of mass. The kinetic energy oscillation amplitudes (shown by blue on the right ordinate) differ by nearly two orders of magnitude. Note that the data for angles less than $24^\circ$ cannot be included due to the lack of any dominant frequency in the amplitude spectrum. Fig.~\ref{Figure_21}(a) shows that the dominant frequency for $\theta=30^\circ$ is found to be $f\approx0.09$. The corresponding values for $\theta=36^\circ$ and $\theta=42^\circ$ are found to be $f\approx0.04$ and $f\approx0.02$ respectively. In order to get an estimate of the frequencies that might be observed in experiments, the dimensionless dominant frequencies reported above need to be converted to dimensional values. For 1mm size particles flowing under the influence of earth's gravity, these frequencies turn out to be approximately $9$ Hz for $\theta=30^\circ$, $4$ Hz for $\theta=36^\circ$ and $2$ Hz for $\theta=42^\circ$. 

\section{\label{sec:Conclusion}Summary and Future work}
We perform two-dimensional DEM simulations of granular materials flowing down a bumpy inclined plane varying over a wide range of inclination angles and calculate various flow properties of interest as the flow evolves with time and achieves a steady state. We also obtain numerical solutions of the momentum balance equations (Eq.~(\ref{x_momentum})) supplemented by different constitutive rheological models (Eqs.~(\ref{mu_i_JFP}) \& (\ref{mu_i_MK})) using the \textit{PDEPE} solver in MATLAB. This solver uses an implicit finite difference method to solve the resulting partial differential equation (Eq.~(\ref{x_momentum_detail})). The numerical simulation assumes a fully developed flow so that it can be compared with the DEM simulations of flow over a chute having periodic boundary conditions.
We first benchmark our numerical solution method with the analytical predictions of \citet{parez2016unsteady} for a linear $\mu-I$ relation. After confirming the accuracy of our numerical solutions, we utilize the non-linear $\mu-I$ relation of the JFP model \cite{jop2006constitutivenature}. The predicted flow properties obtained from the numerical solutions are compared with the DEM simulations and are found to be in good agreement with the DEM simulations for inclination $\theta<28^\circ$. Differences in the predictions of average velocity $v_{avg}$ become noticeable for $\theta=28^\circ$ ($I\sim0.3$) and this difference is attributed to the neglection of normal stress difference in the JFP model. The JFP model fails to capture the flow behavior for $\theta>28^\circ$ as it reaches higher inertial numbers. The deviation between predictions and DEM data keeps increasing with the inclination angle $\theta$ and no steady flow is predicted for $\theta\geq36^\circ$ using the JFP model, which is inconsistent with the DEM results.

Next, we predicted the flow properties using the modified $\mu-I$ rheology along with the normal stress difference law recently proposed by \citet{patro2021rheology}. The method also accounts for the dilation of the layer with an increase in the inertial number as the flow evolves with time.
The predicted flow properties obtained using the modified rheological model are in very good agreement with the DEM simulation results for a much larger range of inclinations (up to $\theta=40^\circ$). In the numerical predictions of the velocity profile, the slip velocity needs to be explicitly accounted for and cannot be neglected at large inclinations. To summarize, we find that accurate prediction of time-dependent flow properties at high inertial numbers requires us to account for the following effects: \newline 
    (1) Non-monotonic $\mu-I$ relation proposed by \citet{patro2021rheology};\newline
    (2) Presence of the normal stress difference;\newline
    (3) Presence of the slip velocity at the chute base; and \newline
    (4) Dilation of the flowing layer to account for changes in the solids fraction.\newline
Ignoring any of these effects leads to significant errors in the predictions at high inclinations. 

Continuum simulations of granular flows using different rheological models are being actively pursued since last few years. However, all such studies have remained limited to inertial numbers less than unity. This work shows that accounting for these four important effects is crucial for a successful and reliable continuum simulation of high-speed granular flows.
It is worth highlighting that the numerical solution of the momentum balance equations does not capture the periodic oscillations in the flow properties at steady state which become prominent at high inclinations. The bulk layer height, the center of mass position of the layer as well as the kinetic energy show oscillations around the steady state value with a characteristic frequency beyond an inclination angle. As the inclination angle increases, both the amplitude and time period of the oscillations increase. Despite the periodic variations observed at high inclinations, the time-averaged properties are predicted reasonably well using the modified rheology proposed by \cite{patro2021rheology}. 

It has been shown by \citet{barker2015well} that the $\mu-I$ rheology given by JFP model is ill-posed at high and low inertial numbers. In other words, for a given set of rheological parameters, there exist a minimum and a maximum value of the inertial number beyond which, the results obtained from solving the rheological and momentum balance equations will be grid size dependent. 
However, the variation of the effective friction coefficient with the inertial number assumed in such studies is not consistent with the simulation results. Hence a detailed study about the well-posedness of the rheological model of \citet{patro2021rheology} by accounting for the presence of the normal stress difference and non-monotonic variation of $\mu$ with $I$ needs to be pursued. Further, the compressibility effects also need to be accounted for. 
In order to check the grid size effect on our predictions, we obtained theoretical predictions using three different grid sizes for the modified rheology case and the results were found to differ from each other by less than $1\%$. Further, the results obtained from the modified rheology are found to be in excellent agreement with the DEM simulations. Hence these results give us confidence that the proposed rheology is able to predict the time-dependent behavior of granular flows at high inertial numbers. 

Most studies using continuum modeling for granular materials treat them as incompressible fluids. Such a treatment for granular flows in silos, hoppers, heaps, etc. may not lead to significant errors since such flows are typically restricted to $I\leq0.5$. However, free surface flows in avalanches and long chutes at high inclinations may show flows at higher inertial numbers. Our results, although reported for 2D, show that ignoring the dilatancy effects in such cases for $I>0.6$ may lead to inaccurate predictions. The presence of normal stress difference reported in the work suggests that the rheology of granular materials in 3D at high inertial numbers will require accounting for both the first and second normal stress differences. A detailed study investigating the study of 3D granular chute flows at high inertial numbers for the transient as well as steady state will be reported in the future. While the solids fraction across the layer remains nearly constant in the case of chute flow, the presence of side walls leads to the variation of the inertial number and hence the solids fraction across the layer in the case of heap flows. Since the modified rheological model of \citet{patro2021rheology} seems to be applicable for both dilute as well as dense granular flow regimes, it can also be utilized to explore the liquid-gas transition that occurs in heap flows confined between two sidewalls at high flow rates \cite{jop2005crucial,louge2005volume}.

\section*{ACKNOWLEDGEMENTS}
AT gratefully acknowledges the financial support provided by the Indian Institute of Technology Kanpur via the initiation grant IITK/CHE/20130338. The authors also acknowledge discussions with Prof. V. Shankar (IIT Kanpur) and Dr. Partha Goswami (IIT Bombay) regarding the characterization of oscillations in the flow.
\section*{DATA AVAILABILITY}
The data that support the findings of this study are available from the corresponding author upon request.
\section*{REFERENCES}
\nocite{*}
\bibliography{aipsamp}%

\providecommand{\noopsort}[1]{}\providecommand{\singleletter}[1]{#1}%
\begin{thebibliography}{76}%
\makeatletter
\providecommand \@ifxundefined [1]{%
 \@ifx{#1\undefined}
}%
\providecommand \@ifnum [1]{%
 \ifnum #1\expandafter \@firstoftwo
 \else \expandafter \@secondoftwo
 \fi
}%
\providecommand \@ifx [1]{%
 \ifx #1\expandafter \@firstoftwo
 \else \expandafter \@secondoftwo
 \fi
}%
\providecommand \natexlab [1]{#1}%
\providecommand \enquote  [1]{``#1''}%
\providecommand \bibnamefont  [1]{#1}%
\providecommand \bibfnamefont [1]{#1}%
\providecommand \citenamefont [1]{#1}%
\providecommand \href@noop [0]{\@secondoftwo}%
\providecommand \href [0]{\begingroup \@sanitize@url \@href}%
\providecommand \@href[1]{\@@startlink{#1}\@@href}%
\providecommand \@@href[1]{\endgroup#1\@@endlink}%
\providecommand \@sanitize@url [0]{\catcode `\\12\catcode `\$12\catcode
  `\&12\catcode `\#12\catcode `\^12\catcode `\_12\catcode `\%12\relax}%
\providecommand \@@startlink[1]{}%
\providecommand \@@endlink[0]{}%
\providecommand \url  [0]{\begingroup\@sanitize@url \@url }%
\providecommand \@url [1]{\endgroup\@href {#1}{\urlprefix }}%
\providecommand \urlprefix  [0]{URL }%
\providecommand \Eprint [0]{\href }%
\providecommand \doibase [0]{https://doi.org/}%
\providecommand \selectlanguage [0]{\@gobble}%
\providecommand \bibinfo  [0]{\@secondoftwo}%
\providecommand \bibfield  [0]{\@secondoftwo}%
\providecommand \translation [1]{[#1]}%
\providecommand \BibitemOpen [0]{}%
\providecommand \bibitemStop [0]{}%
\providecommand \bibitemNoStop [0]{.\EOS\space}%
\providecommand \EOS [0]{\spacefactor3000\relax}%
\providecommand \BibitemShut  [1]{\csname bibitem#1\endcsname}%
\let\auto@bib@innerbib\@empty
\bibitem [{\citenamefont
  {Pouliquen}(1999{\natexlab{a}})}]{pouliquen1999scaling}%
  \BibitemOpen
  \bibfield  {author} {\bibinfo {author} {\bibfnamefont {O.}~\bibnamefont
  {Pouliquen}},\ }\bibfield  {title} {\bibinfo {title} {Scaling laws in
  granular flows down rough inclined planes},\ }\href@noop {} {\bibfield
  {journal} {\bibinfo  {journal} {Phys. Fluids}\ }\textbf {\bibinfo {volume}
  {11}},\ \bibinfo {pages} {542} (\bibinfo {year}
  {1999}{\natexlab{a}})}\BibitemShut {NoStop}%
\bibitem [{\citenamefont {MiDi}(2004)}]{midi2004dense}%
  \BibitemOpen
  \bibfield  {author} {\bibinfo {author} {\bibfnamefont {G.}~\bibnamefont
  {MiDi}},\ }\bibfield  {title} {\bibinfo {title} {On dense granular flows.},\
  }\href@noop {} {\bibfield  {journal} {\bibinfo  {journal} {European Physical
  Journal E--Soft Matter}\ }\textbf {\bibinfo {volume} {14}} (\bibinfo {year}
  {2004})}\BibitemShut {NoStop}%
\bibitem [{\citenamefont {Jop}\ \emph {et~al.}(2005)\citenamefont {Jop},
  \citenamefont {Forterre},\ and\ \citenamefont {Pouliquen}}]{jop2005crucial}%
  \BibitemOpen
  \bibfield  {author} {\bibinfo {author} {\bibfnamefont {P.}~\bibnamefont
  {Jop}}, \bibinfo {author} {\bibfnamefont {Y.}~\bibnamefont {Forterre}},\ and\
  \bibinfo {author} {\bibfnamefont {O.}~\bibnamefont {Pouliquen}},\ }\bibfield
  {title} {\bibinfo {title} {Crucial role of sidewalls in granular surface
  flows: consequences for the rheology},\ }\href@noop {} {\bibfield  {journal}
  {\bibinfo  {journal} {J. Fluid Mech.}\ }\textbf {\bibinfo {volume} {541}},\
  \bibinfo {pages} {167} (\bibinfo {year} {2005})}\BibitemShut {NoStop}%
\bibitem [{\citenamefont {Holyoake}\ and\ \citenamefont
  {McElwaine}(2012)}]{holyoake2012high}%
  \BibitemOpen
  \bibfield  {author} {\bibinfo {author} {\bibfnamefont {A.~J.}\ \bibnamefont
  {Holyoake}}\ and\ \bibinfo {author} {\bibfnamefont {J.~N.}\ \bibnamefont
  {McElwaine}},\ }\bibfield  {title} {\bibinfo {title} {High-speed granular
  chute flows},\ }\href@noop {} {\bibfield  {journal} {\bibinfo  {journal} {J.
  Fluid Mech.}\ }\textbf {\bibinfo {volume} {710}},\ \bibinfo {pages} {35}
  (\bibinfo {year} {2012})}\BibitemShut {NoStop}%
\bibitem [{\citenamefont {Pouliquen}(1999{\natexlab{b}})}]{pouliquen1999shape}%
  \BibitemOpen
  \bibfield  {author} {\bibinfo {author} {\bibfnamefont {O.}~\bibnamefont
  {Pouliquen}},\ }\bibfield  {title} {\bibinfo {title} {On the shape of
  granular fronts down rough inclined planes},\ }\href@noop {} {\bibfield
  {journal} {\bibinfo  {journal} {Phys. Fluids}\ }\textbf {\bibinfo {volume}
  {11}},\ \bibinfo {pages} {1956} (\bibinfo {year}
  {1999}{\natexlab{b}})}\BibitemShut {NoStop}%
\bibitem [{\citenamefont {Heyman}\ \emph
  {et~al.}(2017{\natexlab{a}})\citenamefont {Heyman}, \citenamefont
  {Boltenhagen}, \citenamefont {Delannay},\ and\ \citenamefont
  {Valance}}]{heyman2017experimental}%
  \BibitemOpen
  \bibfield  {author} {\bibinfo {author} {\bibfnamefont {J.}~\bibnamefont
  {Heyman}}, \bibinfo {author} {\bibfnamefont {P.}~\bibnamefont {Boltenhagen}},
  \bibinfo {author} {\bibfnamefont {R.}~\bibnamefont {Delannay}},\ and\
  \bibinfo {author} {\bibfnamefont {A.}~\bibnamefont {Valance}},\ }\bibfield
  {title} {\bibinfo {title} {Experimental investigation of high speed granular
  flows down inclines},\ }in\ \href@noop {} {\emph {\bibinfo {booktitle} {EPJ
  Web of Conferences}}},\ Vol.\ \bibinfo {volume} {140}\ (\bibinfo {year}
  {2017})\ p.\ \bibinfo {pages} {03057}\BibitemShut {NoStop}%
\bibitem [{\citenamefont {Thompson}\ and\ \citenamefont
  {Huppert}(2007)}]{thompson2007granular}%
  \BibitemOpen
  \bibfield  {author} {\bibinfo {author} {\bibfnamefont {E.~L.}\ \bibnamefont
  {Thompson}}\ and\ \bibinfo {author} {\bibfnamefont {H.~E.}\ \bibnamefont
  {Huppert}},\ }\bibfield  {title} {\bibinfo {title} {Granular column
  collapses: further experimental results},\ }\href@noop {} {\bibfield
  {journal} {\bibinfo  {journal} {J. Fluid Mech.}\ }\textbf {\bibinfo {volume}
  {575}},\ \bibinfo {pages} {177} (\bibinfo {year} {2007})}\BibitemShut
  {NoStop}%
\bibitem [{\citenamefont {Lacaze}\ \emph {et~al.}(2008)\citenamefont {Lacaze},
  \citenamefont {Phillips},\ and\ \citenamefont {Kerswell}}]{lacaze2008planar}%
  \BibitemOpen
  \bibfield  {author} {\bibinfo {author} {\bibfnamefont {L.}~\bibnamefont
  {Lacaze}}, \bibinfo {author} {\bibfnamefont {J.~C.}\ \bibnamefont
  {Phillips}},\ and\ \bibinfo {author} {\bibfnamefont {R.~R.}\ \bibnamefont
  {Kerswell}},\ }\bibfield  {title} {\bibinfo {title} {Planar collapse of a
  granular column: Experiments and discrete element simulations},\ }\href@noop
  {} {\bibfield  {journal} {\bibinfo  {journal} {Phys. Fluids}\ }\textbf
  {\bibinfo {volume} {20}},\ \bibinfo {pages} {063302} (\bibinfo {year}
  {2008})}\BibitemShut {NoStop}%
\bibitem [{\citenamefont {Lube}\ \emph {et~al.}(2004)\citenamefont {Lube},
  \citenamefont {Huppert}, \citenamefont {Sparks},\ and\ \citenamefont
  {Hallworth}}]{lube2004axisymmetric}%
  \BibitemOpen
  \bibfield  {author} {\bibinfo {author} {\bibfnamefont {G.}~\bibnamefont
  {Lube}}, \bibinfo {author} {\bibfnamefont {H.~E.}\ \bibnamefont {Huppert}},
  \bibinfo {author} {\bibfnamefont {R.~S.~J.}\ \bibnamefont {Sparks}},\ and\
  \bibinfo {author} {\bibfnamefont {M.~A.}\ \bibnamefont {Hallworth}},\
  }\bibfield  {title} {\bibinfo {title} {Axisymmetric collapses of granular
  columns},\ }\href@noop {} {\bibfield  {journal} {\bibinfo  {journal} {J.
  Fluid Mech.}\ }\textbf {\bibinfo {volume} {508}},\ \bibinfo {pages} {175}
  (\bibinfo {year} {2004})}\BibitemShut {NoStop}%
\bibitem [{\citenamefont {Lajeunesse}\ \emph {et~al.}(2004)\citenamefont
  {Lajeunesse}, \citenamefont {Mangeney-Castelnau},\ and\ \citenamefont
  {Vilotte}}]{lajeunesse2004spreading}%
  \BibitemOpen
  \bibfield  {author} {\bibinfo {author} {\bibfnamefont {E.}~\bibnamefont
  {Lajeunesse}}, \bibinfo {author} {\bibfnamefont {A.}~\bibnamefont
  {Mangeney-Castelnau}},\ and\ \bibinfo {author} {\bibfnamefont
  {J.}~\bibnamefont {Vilotte}},\ }\bibfield  {title} {\bibinfo {title}
  {Spreading of a granular mass on a horizontal plane},\ }\href@noop {}
  {\bibfield  {journal} {\bibinfo  {journal} {Phys. Fluids}\ }\textbf {\bibinfo
  {volume} {16}},\ \bibinfo {pages} {2371} (\bibinfo {year}
  {2004})}\BibitemShut {NoStop}%
\bibitem [{\citenamefont {Zhou}\ \emph {et~al.}(2017)\citenamefont {Zhou},
  \citenamefont {Lagr{\'e}e}, \citenamefont {Popinet}, \citenamefont {Ruyer},\
  and\ \citenamefont {Aussillous}}]{zhou2017experiments}%
  \BibitemOpen
  \bibfield  {author} {\bibinfo {author} {\bibfnamefont {Y.}~\bibnamefont
  {Zhou}}, \bibinfo {author} {\bibfnamefont {P.-Y.}\ \bibnamefont
  {Lagr{\'e}e}}, \bibinfo {author} {\bibfnamefont {S.}~\bibnamefont {Popinet}},
  \bibinfo {author} {\bibfnamefont {P.}~\bibnamefont {Ruyer}},\ and\ \bibinfo
  {author} {\bibfnamefont {P.}~\bibnamefont {Aussillous}},\ }\bibfield  {title}
  {\bibinfo {title} {Experiments on, and discrete and continuum simulations of,
  the discharge of granular media from silos with a lateral orifice},\
  }\href@noop {} {\bibfield  {journal} {\bibinfo  {journal} {J. Fluid Mech.}\
  }\textbf {\bibinfo {volume} {829}},\ \bibinfo {pages} {459} (\bibinfo {year}
  {2017})}\BibitemShut {NoStop}%
\bibitem [{\citenamefont {Khakhar}\ \emph {et~al.}(1997)\citenamefont
  {Khakhar}, \citenamefont {McCarthy},\ and\ \citenamefont
  {Ottino}}]{khakhar1997radial}%
  \BibitemOpen
  \bibfield  {author} {\bibinfo {author} {\bibfnamefont {D.}~\bibnamefont
  {Khakhar}}, \bibinfo {author} {\bibfnamefont {J.}~\bibnamefont {McCarthy}},\
  and\ \bibinfo {author} {\bibfnamefont {J.~M.}\ \bibnamefont {Ottino}},\
  }\bibfield  {title} {\bibinfo {title} {Radial segregation of granular
  mixtures in rotating cylinders},\ }\href@noop {} {\bibfield  {journal}
  {\bibinfo  {journal} {Phys. Fluids}\ }\textbf {\bibinfo {volume} {9}},\
  \bibinfo {pages} {3600} (\bibinfo {year} {1997})}\BibitemShut {NoStop}%
\bibitem [{\citenamefont {Ji}\ \emph {et~al.}(2009)\citenamefont {Ji},
  \citenamefont {Hanes},\ and\ \citenamefont
  {Shen}}]{annular_shear_cell_2009comparisons}%
  \BibitemOpen
  \bibfield  {author} {\bibinfo {author} {\bibfnamefont {S.}~\bibnamefont
  {Ji}}, \bibinfo {author} {\bibfnamefont {D.~M.}\ \bibnamefont {Hanes}},\ and\
  \bibinfo {author} {\bibfnamefont {H.~H.}\ \bibnamefont {Shen}},\ }\bibfield
  {title} {\bibinfo {title} {Comparisons of physical experiment and discrete
  element simulations of sheared granular materials in an annular shear cell},\
  }\href@noop {} {\bibfield  {journal} {\bibinfo  {journal} {Mechanics of
  materials}\ }\textbf {\bibinfo {volume} {41}},\ \bibinfo {pages} {764}
  (\bibinfo {year} {2009})}\BibitemShut {NoStop}%
\bibitem [{\citenamefont {Azanza}\ \emph {et~al.}(1999)\citenamefont {Azanza},
  \citenamefont {Chevoir},\ and\ \citenamefont
  {Moucheront}}]{azanza1999experimental}%
  \BibitemOpen
  \bibfield  {author} {\bibinfo {author} {\bibfnamefont {E.}~\bibnamefont
  {Azanza}}, \bibinfo {author} {\bibfnamefont {F.}~\bibnamefont {Chevoir}},\
  and\ \bibinfo {author} {\bibfnamefont {P.}~\bibnamefont {Moucheront}},\
  }\bibfield  {title} {\bibinfo {title} {Experimental study of collisional
  granular flows down an inclined plane},\ }\href@noop {} {\bibfield  {journal}
  {\bibinfo  {journal} {J. Fluid Mech.}\ }\textbf {\bibinfo {volume} {400}},\
  \bibinfo {pages} {199} (\bibinfo {year} {1999})}\BibitemShut {NoStop}%
\bibitem [{\citenamefont {Yang}\ and\ \citenamefont
  {Hsiau}(2001)}]{yang2001simulation}%
  \BibitemOpen
  \bibfield  {author} {\bibinfo {author} {\bibfnamefont {S.-C.}\ \bibnamefont
  {Yang}}\ and\ \bibinfo {author} {\bibfnamefont {S.-S.}\ \bibnamefont
  {Hsiau}},\ }\bibfield  {title} {\bibinfo {title} {The simulation and
  experimental study of granular materials discharged from a silo with the
  placement of inserts},\ }\href@noop {} {\bibfield  {journal} {\bibinfo
  {journal} {Powder Technology}\ }\textbf {\bibinfo {volume} {120}},\ \bibinfo
  {pages} {244} (\bibinfo {year} {2001})}\BibitemShut {NoStop}%
\bibitem [{\citenamefont {Orpe}\ and\ \citenamefont
  {Khakhar}(2007)}]{orpe2007rheology}%
  \BibitemOpen
  \bibfield  {author} {\bibinfo {author} {\bibfnamefont {A.~V.}\ \bibnamefont
  {Orpe}}\ and\ \bibinfo {author} {\bibfnamefont {D.}~\bibnamefont {Khakhar}},\
  }\bibfield  {title} {\bibinfo {title} {Rheology of surface granular flows},\
  }\href@noop {} {\bibfield  {journal} {\bibinfo  {journal} {J. Fluid Mech.}\
  }\textbf {\bibinfo {volume} {571}},\ \bibinfo {pages} {1} (\bibinfo {year}
  {2007})}\BibitemShut {NoStop}%
\bibitem [{\citenamefont {Silbert}\ \emph {et~al.}(2001)\citenamefont
  {Silbert}, \citenamefont {Erta{\c{s}}}, \citenamefont {Grest}, \citenamefont
  {Halsey}, \citenamefont {Levine},\ and\ \citenamefont
  {Plimpton}}]{silbert2001granular}%
  \BibitemOpen
  \bibfield  {author} {\bibinfo {author} {\bibfnamefont {L.~E.}\ \bibnamefont
  {Silbert}}, \bibinfo {author} {\bibfnamefont {D.}~\bibnamefont
  {Erta{\c{s}}}}, \bibinfo {author} {\bibfnamefont {G.~S.}\ \bibnamefont
  {Grest}}, \bibinfo {author} {\bibfnamefont {T.~C.}\ \bibnamefont {Halsey}},
  \bibinfo {author} {\bibfnamefont {D.}~\bibnamefont {Levine}},\ and\ \bibinfo
  {author} {\bibfnamefont {S.~J.}\ \bibnamefont {Plimpton}},\ }\bibfield
  {title} {\bibinfo {title} {Granular flow down an inclined plane: Bagnold
  scaling and rheology},\ }\href@noop {} {\bibfield  {journal} {\bibinfo
  {journal} {Phys. Rev. E}\ }\textbf {\bibinfo {volume} {64}},\ \bibinfo
  {pages} {051302} (\bibinfo {year} {2001})}\BibitemShut {NoStop}%
\bibitem [{\citenamefont {Da~Cruz}\ \emph {et~al.}(2005)\citenamefont
  {Da~Cruz}, \citenamefont {Emam}, \citenamefont {Prochnow}, \citenamefont
  {Roux},\ and\ \citenamefont {Chevoir}}]{da2005rheophysics}%
  \BibitemOpen
  \bibfield  {author} {\bibinfo {author} {\bibfnamefont {F.}~\bibnamefont
  {Da~Cruz}}, \bibinfo {author} {\bibfnamefont {S.}~\bibnamefont {Emam}},
  \bibinfo {author} {\bibfnamefont {M.}~\bibnamefont {Prochnow}}, \bibinfo
  {author} {\bibfnamefont {J.-N.}\ \bibnamefont {Roux}},\ and\ \bibinfo
  {author} {\bibfnamefont {F.}~\bibnamefont {Chevoir}},\ }\bibfield  {title}
  {\bibinfo {title} {Rheophysics of dense granular materials: Discrete
  simulation of plane shear flows},\ }\href@noop {} {\bibfield  {journal}
  {\bibinfo  {journal} {Phys. Rev. E}\ }\textbf {\bibinfo {volume} {72}},\
  \bibinfo {pages} {021309} (\bibinfo {year} {2005})}\BibitemShut {NoStop}%
\bibitem [{\citenamefont {Pouliquen}\ \emph {et~al.}(2006)\citenamefont
  {Pouliquen}, \citenamefont {Cassar}, \citenamefont {Jop}, \citenamefont
  {Forterre},\ and\ \citenamefont {Nicolas}}]{pouliquen2006flow}%
  \BibitemOpen
  \bibfield  {author} {\bibinfo {author} {\bibfnamefont {O.}~\bibnamefont
  {Pouliquen}}, \bibinfo {author} {\bibfnamefont {C.}~\bibnamefont {Cassar}},
  \bibinfo {author} {\bibfnamefont {P.}~\bibnamefont {Jop}}, \bibinfo {author}
  {\bibfnamefont {Y.}~\bibnamefont {Forterre}},\ and\ \bibinfo {author}
  {\bibfnamefont {M.}~\bibnamefont {Nicolas}},\ }\bibfield  {title} {\bibinfo
  {title} {Flow of dense granular material: towards simple constitutive laws},\
  }\href@noop {} {\bibfield  {journal} {\bibinfo  {journal} {Journal of
  Statistical Mechanics: Theory and Experiment}\ }\textbf {\bibinfo {volume}
  {2006}},\ \bibinfo {pages} {P07020} (\bibinfo {year} {2006})}\BibitemShut
  {NoStop}%
\bibitem [{\citenamefont {Baran}\ \emph {et~al.}(2006)\citenamefont {Baran},
  \citenamefont {Erta\ifmmode~\mbox{\c{s}}\else \c{s}\fi{}}, \citenamefont
  {Halsey}, \citenamefont {Grest},\ and\ \citenamefont {Lechman}}]{baran_2006}%
  \BibitemOpen
  \bibfield  {author} {\bibinfo {author} {\bibfnamefont {O.}~\bibnamefont
  {Baran}}, \bibinfo {author} {\bibfnamefont {D.}~\bibnamefont
  {Erta\ifmmode~\mbox{\c{s}}\else \c{s}\fi{}}}, \bibinfo {author}
  {\bibfnamefont {T.~C.}\ \bibnamefont {Halsey}}, \bibinfo {author}
  {\bibfnamefont {G.~S.}\ \bibnamefont {Grest}},\ and\ \bibinfo {author}
  {\bibfnamefont {J.~B.}\ \bibnamefont {Lechman}},\ }\bibfield  {title}
  {\bibinfo {title} {Velocity correlations in dense gravity-driven granular
  chute flow},\ }\href {https://doi.org/10.1103/PhysRevE.74.051302} {\bibfield
  {journal} {\bibinfo  {journal} {Phys. Rev. E}\ }\textbf {\bibinfo {volume}
  {74}},\ \bibinfo {pages} {051302} (\bibinfo {year} {2006})}\BibitemShut
  {NoStop}%
\bibitem [{\citenamefont {B{\"o}rzs{\"o}nyi}\ \emph {et~al.}(2009)\citenamefont
  {B{\"o}rzs{\"o}nyi}, \citenamefont {Ecke},\ and\ \citenamefont
  {McElwaine}}]{borzsonyi2009patterns}%
  \BibitemOpen
  \bibfield  {author} {\bibinfo {author} {\bibfnamefont {T.}~\bibnamefont
  {B{\"o}rzs{\"o}nyi}}, \bibinfo {author} {\bibfnamefont {R.~E.}\ \bibnamefont
  {Ecke}},\ and\ \bibinfo {author} {\bibfnamefont {J.~N.}\ \bibnamefont
  {McElwaine}},\ }\bibfield  {title} {\bibinfo {title} {Patterns in flowing
  sand: understanding the physics of granular flow},\ }\href@noop {} {\bibfield
   {journal} {\bibinfo  {journal} {Phys. Rev. Lett.}\ }\textbf {\bibinfo
  {volume} {103}},\ \bibinfo {pages} {178302} (\bibinfo {year}
  {2009})}\BibitemShut {NoStop}%
\bibitem [{\citenamefont {Tripathi}\ and\ \citenamefont
  {Khakhar}(2011)}]{tripathi2011rheology}%
  \BibitemOpen
  \bibfield  {author} {\bibinfo {author} {\bibfnamefont {A.}~\bibnamefont
  {Tripathi}}\ and\ \bibinfo {author} {\bibfnamefont {D.}~\bibnamefont
  {Khakhar}},\ }\bibfield  {title} {\bibinfo {title} {Rheology of binary
  granular mixtures in the dense flow regime},\ }\href@noop {} {\bibfield
  {journal} {\bibinfo  {journal} {Phys. Fluids}\ }\textbf {\bibinfo {volume}
  {23}},\ \bibinfo {pages} {113302} (\bibinfo {year} {2011})}\BibitemShut
  {NoStop}%
\bibitem [{\citenamefont {Kumaran}\ and\ \citenamefont
  {Bharathraj}(2013)}]{kumaran2013}%
  \BibitemOpen
  \bibfield  {author} {\bibinfo {author} {\bibfnamefont {V.}~\bibnamefont
  {Kumaran}}\ and\ \bibinfo {author} {\bibfnamefont {S.}~\bibnamefont
  {Bharathraj}},\ }\bibfield  {title} {\bibinfo {title} {The effect of base
  roughness on the development of a dense granular flow down an inclined
  plane},\ }\href@noop {} {\bibfield  {journal} {\bibinfo  {journal} {Phys.
  Fluids}\ }\textbf {\bibinfo {volume} {25}},\ \bibinfo {pages} {070604}
  (\bibinfo {year} {2013})}\BibitemShut {NoStop}%
\bibitem [{\citenamefont {Brodu}\ \emph {et~al.}(2015)\citenamefont {Brodu},
  \citenamefont {Delannay}, \citenamefont {Valance},\ and\ \citenamefont
  {Richard}}]{brodu_delannay_valance_richard_2015}%
  \BibitemOpen
  \bibfield  {author} {\bibinfo {author} {\bibfnamefont {N.}~\bibnamefont
  {Brodu}}, \bibinfo {author} {\bibfnamefont {R.}~\bibnamefont {Delannay}},
  \bibinfo {author} {\bibfnamefont {A.}~\bibnamefont {Valance}},\ and\ \bibinfo
  {author} {\bibfnamefont {P.}~\bibnamefont {Richard}},\ }\bibfield  {title}
  {\bibinfo {title} {New patterns in high-speed granular flows},\ }\href
  {https://doi.org/10.1017/jfm.2015.109} {\bibfield  {journal} {\bibinfo
  {journal} {J. Fluid Mech.}\ }\textbf {\bibinfo {volume} {769}},\ \bibinfo
  {pages} {218–228} (\bibinfo {year} {2015})}\BibitemShut {NoStop}%
\bibitem [{\citenamefont {Mandal}\ and\ \citenamefont
  {Khakhar}(2016)}]{mandal2016study}%
  \BibitemOpen
  \bibfield  {author} {\bibinfo {author} {\bibfnamefont {S.}~\bibnamefont
  {Mandal}}\ and\ \bibinfo {author} {\bibfnamefont {D.~V.}\ \bibnamefont
  {Khakhar}},\ }\bibfield  {title} {\bibinfo {title} {A study of the rheology
  of planar granular flow of dumbbells using discrete element method
  simulations},\ }\href@noop {} {\bibfield  {journal} {\bibinfo  {journal}
  {Phys. Fluids}\ }\textbf {\bibinfo {volume} {28}},\ \bibinfo {pages} {103301}
  (\bibinfo {year} {2016})}\BibitemShut {NoStop}%
\bibitem [{\citenamefont {Ralaiarisoa}\ \emph {et~al.}(2017)\citenamefont
  {Ralaiarisoa}, \citenamefont {Valance}, \citenamefont {Brodu},\ and\
  \citenamefont {Delannay}}]{ralaiarisoa2017high}%
  \BibitemOpen
  \bibfield  {author} {\bibinfo {author} {\bibfnamefont {V.~J.-L.}\
  \bibnamefont {Ralaiarisoa}}, \bibinfo {author} {\bibfnamefont
  {A.}~\bibnamefont {Valance}}, \bibinfo {author} {\bibfnamefont
  {N.}~\bibnamefont {Brodu}},\ and\ \bibinfo {author} {\bibfnamefont
  {R.}~\bibnamefont {Delannay}},\ }\bibfield  {title} {\bibinfo {title} {High
  speed confined granular flows down inclined: numerical simulations},\ }in\
  \href@noop {} {\emph {\bibinfo {booktitle} {EPJ Web of Conferences}}},\ Vol.\
  \bibinfo {volume} {140}\ (\bibinfo {year} {2017})\ p.\ \bibinfo {pages}
  {03081}\BibitemShut {NoStop}%
\bibitem [{\citenamefont {Mandal}\ and\ \citenamefont
  {Khakhar}(2018)}]{mandal2018study}%
  \BibitemOpen
  \bibfield  {author} {\bibinfo {author} {\bibfnamefont {S.}~\bibnamefont
  {Mandal}}\ and\ \bibinfo {author} {\bibfnamefont {D.}~\bibnamefont
  {Khakhar}},\ }\bibfield  {title} {\bibinfo {title} {A study of the rheology
  and micro-structure of dumbbells in shear geometries},\ }\href@noop {}
  {\bibfield  {journal} {\bibinfo  {journal} {Phys. Fluids}\ }\textbf {\bibinfo
  {volume} {30}},\ \bibinfo {pages} {013303} (\bibinfo {year}
  {2018})}\BibitemShut {NoStop}%
\bibitem [{\citenamefont {Bhateja}\ and\ \citenamefont
  {Khakhar}(2018)}]{bhateja2018}%
  \BibitemOpen
  \bibfield  {author} {\bibinfo {author} {\bibfnamefont {A.}~\bibnamefont
  {Bhateja}}\ and\ \bibinfo {author} {\bibfnamefont {D.~V.}\ \bibnamefont
  {Khakhar}},\ }\bibfield  {title} {\bibinfo {title} {Rheology of dense
  granular flows in two dimensions: Comparison of fully two-dimensional flows
  to unidirectional shear flow},\ }\href@noop {} {\bibfield  {journal}
  {\bibinfo  {journal} {Phys. Rev. Fluids}\ }\textbf {\bibinfo {volume} {3}},\
  \bibinfo {pages} {062301} (\bibinfo {year} {2018})}\BibitemShut {NoStop}%
\bibitem [{\citenamefont {Bhateja}\ and\ \citenamefont
  {Khakhar}(2020)}]{bhateja2020}%
  \BibitemOpen
  \bibfield  {author} {\bibinfo {author} {\bibfnamefont {A.}~\bibnamefont
  {Bhateja}}\ and\ \bibinfo {author} {\bibfnamefont {D.~V.}\ \bibnamefont
  {Khakhar}},\ }\bibfield  {title} {\bibinfo {title} {Analysis of granular
  rheology in a quasi-two-dimensional slow flow by means of discrete element
  method based simulations},\ }\href@noop {} {\bibfield  {journal} {\bibinfo
  {journal} {Phys. Fluids}\ }\textbf {\bibinfo {volume} {32}},\ \bibinfo
  {pages} {013301} (\bibinfo {year} {2020})}\BibitemShut {NoStop}%
\bibitem [{\citenamefont {Patro}\ \emph {et~al.}(2021)\citenamefont {Patro},
  \citenamefont {Prasad}, \citenamefont {Tripathi}, \citenamefont {Kumar},\
  and\ \citenamefont {Tripathi}}]{patro2021rheology}%
  \BibitemOpen
  \bibfield  {author} {\bibinfo {author} {\bibfnamefont {S.}~\bibnamefont
  {Patro}}, \bibinfo {author} {\bibfnamefont {M.}~\bibnamefont {Prasad}},
  \bibinfo {author} {\bibfnamefont {A.}~\bibnamefont {Tripathi}}, \bibinfo
  {author} {\bibfnamefont {P.}~\bibnamefont {Kumar}},\ and\ \bibinfo {author}
  {\bibfnamefont {A.}~\bibnamefont {Tripathi}},\ }\bibfield  {title} {\bibinfo
  {title} {Rheology of two-dimensional granular chute flows at high inertial
  numbers},\ }\href@noop {} {\bibfield  {journal} {\bibinfo  {journal} {Phys.
  Fluids}\ }\textbf {\bibinfo {volume} {33}},\ \bibinfo {pages} {113321}
  (\bibinfo {year} {2021})}\BibitemShut {NoStop}%
\bibitem [{\citenamefont {Debnath}\ \emph
  {et~al.}(2022{\natexlab{a}})\citenamefont {Debnath}, \citenamefont {Rao},\
  and\ \citenamefont {Kumaran}}]{debnath2022different}%
  \BibitemOpen
  \bibfield  {author} {\bibinfo {author} {\bibfnamefont {B.}~\bibnamefont
  {Debnath}}, \bibinfo {author} {\bibfnamefont {K.~K.}\ \bibnamefont {Rao}},\
  and\ \bibinfo {author} {\bibfnamefont {V.}~\bibnamefont {Kumaran}},\
  }\bibfield  {title} {\bibinfo {title} {Different shear regimes in the dense
  granular flow in a vertical channel},\ }\href@noop {} {\bibfield  {journal}
  {\bibinfo  {journal} {J. Fluid Mech.}\ }\textbf {\bibinfo {volume} {945}},\
  \bibinfo {pages} {A25} (\bibinfo {year} {2022}{\natexlab{a}})}\BibitemShut
  {NoStop}%
\bibitem [{\citenamefont {Forterre}\ and\ \citenamefont
  {Pouliquen}(2008)}]{forterre2008flows}%
  \BibitemOpen
  \bibfield  {author} {\bibinfo {author} {\bibfnamefont {Y.}~\bibnamefont
  {Forterre}}\ and\ \bibinfo {author} {\bibfnamefont {O.}~\bibnamefont
  {Pouliquen}},\ }\bibfield  {title} {\bibinfo {title} {Flows of dense granular
  media},\ }\href@noop {} {\bibfield  {journal} {\bibinfo  {journal} {Annu.
  Rev. Fluid Mech.}\ }\textbf {\bibinfo {volume} {40}},\ \bibinfo {pages} {1}
  (\bibinfo {year} {2008})}\BibitemShut {NoStop}%
\bibitem [{\citenamefont {B.~Andreotti}\ and\ \citenamefont
  {Pouliquen}(2013)}]{andereotti_forterre_pouliquen_2013}%
  \BibitemOpen
  \bibfield  {author} {\bibinfo {author} {\bibfnamefont {Y.~F.}\ \bibnamefont
  {B.~Andreotti}}\ and\ \bibinfo {author} {\bibfnamefont {O.}~\bibnamefont
  {Pouliquen}},\ }\href@noop {} {\emph {\bibinfo {title} {Granular Media:
  Between Fluid and Solid}}}\ (\bibinfo  {publisher} {Cambridge University
  Press},\ \bibinfo {address} {Cambridge},\ \bibinfo {year} {2013})\BibitemShut
  {NoStop}%
\bibitem [{\citenamefont {Jop}(2015)}]{jop2015rheological}%
  \BibitemOpen
  \bibfield  {author} {\bibinfo {author} {\bibfnamefont {P.}~\bibnamefont
  {Jop}},\ }\bibfield  {title} {\bibinfo {title} {Rheological properties of
  dense granular flows},\ }\href@noop {} {\bibfield  {journal} {\bibinfo
  {journal} {Comptes rendus physique}\ }\textbf {\bibinfo {volume} {16}},\
  \bibinfo {pages} {62} (\bibinfo {year} {2015})}\BibitemShut {NoStop}%
\bibitem [{\citenamefont {Jop}\ \emph {et~al.}(2006)\citenamefont {Jop},
  \citenamefont {Forterre},\ and\ \citenamefont
  {Pouliquen}}]{jop2006constitutivenature}%
  \BibitemOpen
  \bibfield  {author} {\bibinfo {author} {\bibfnamefont {P.}~\bibnamefont
  {Jop}}, \bibinfo {author} {\bibfnamefont {Y.}~\bibnamefont {Forterre}},\ and\
  \bibinfo {author} {\bibfnamefont {O.}~\bibnamefont {Pouliquen}},\ }\bibfield
  {title} {\bibinfo {title} {A constitutive law for dense granular flows},\
  }\href@noop {} {\bibfield  {journal} {\bibinfo  {journal} {Nature}\ }\textbf
  {\bibinfo {volume} {441}},\ \bibinfo {pages} {727} (\bibinfo {year}
  {2006})}\BibitemShut {NoStop}%
\bibitem [{\citenamefont {Lagrée}\ \emph {et~al.}(2011)\citenamefont
  {Lagrée}, \citenamefont {Staron},\ and\ \citenamefont
  {Popinet}}]{lagree_staron_popinet_2011}%
  \BibitemOpen
  \bibfield  {author} {\bibinfo {author} {\bibfnamefont {P.-Y.}\ \bibnamefont
  {Lagrée}}, \bibinfo {author} {\bibfnamefont {L.}~\bibnamefont {Staron}},\
  and\ \bibinfo {author} {\bibfnamefont {S.}~\bibnamefont {Popinet}},\
  }\bibfield  {title} {\bibinfo {title} {The granular column collapse as a
  continuum: validity of a two-dimensional navier–stokes model with a
  $\mu(\mathrm{I})$-rheology},\ }\href {https://doi.org/10.1017/jfm.2011.335}
  {\bibfield  {journal} {\bibinfo  {journal} {J. Fluid Mech.}\ }\textbf
  {\bibinfo {volume} {686}},\ \bibinfo {pages} {378–408} (\bibinfo {year}
  {2011})}\BibitemShut {NoStop}%
\bibitem [{\citenamefont {Renouf}\ \emph {et~al.}(2005)\citenamefont {Renouf},
  \citenamefont {Bonamy}, \citenamefont {Dubois},\ and\ \citenamefont
  {Alart}}]{renouf2005numerical}%
  \BibitemOpen
  \bibfield  {author} {\bibinfo {author} {\bibfnamefont {M.}~\bibnamefont
  {Renouf}}, \bibinfo {author} {\bibfnamefont {D.}~\bibnamefont {Bonamy}},
  \bibinfo {author} {\bibfnamefont {F.}~\bibnamefont {Dubois}},\ and\ \bibinfo
  {author} {\bibfnamefont {P.}~\bibnamefont {Alart}},\ }\bibfield  {title}
  {\bibinfo {title} {Numerical simulation of two-dimensional steady granular
  flows in rotating drum: On surface flow rheology},\ }\href@noop {} {\bibfield
   {journal} {\bibinfo  {journal} {Phys. Fluids}\ }\textbf {\bibinfo {volume}
  {17}},\ \bibinfo {pages} {103303} (\bibinfo {year} {2005})}\BibitemShut
  {NoStop}%
\bibitem [{\citenamefont {Forterre}\ and\ \citenamefont
  {Pouliquen}(2003)}]{forterre2003long}%
  \BibitemOpen
  \bibfield  {author} {\bibinfo {author} {\bibfnamefont {Y.}~\bibnamefont
  {Forterre}}\ and\ \bibinfo {author} {\bibfnamefont {O.}~\bibnamefont
  {Pouliquen}},\ }\bibfield  {title} {\bibinfo {title} {Long-surface-wave
  instability in dense granular flows},\ }\href@noop {} {\bibfield  {journal}
  {\bibinfo  {journal} {J. Fluid Mech.}\ }\textbf {\bibinfo {volume} {486}},\
  \bibinfo {pages} {21} (\bibinfo {year} {2003})}\BibitemShut {NoStop}%
\bibitem [{\citenamefont {Mandal}\ and\ \citenamefont
  {Khakhar}(2017)}]{mandal2017sidewall}%
  \BibitemOpen
  \bibfield  {author} {\bibinfo {author} {\bibfnamefont {S.}~\bibnamefont
  {Mandal}}\ and\ \bibinfo {author} {\bibfnamefont {D.}~\bibnamefont
  {Khakhar}},\ }\bibfield  {title} {\bibinfo {title} {Sidewall-friction-driven
  ordering transition in granular channel flows: Implications for granular
  rheology},\ }\href@noop {} {\bibfield  {journal} {\bibinfo  {journal} {Phys.
  Rev. E}\ }\textbf {\bibinfo {volume} {96}},\ \bibinfo {pages} {050901}
  (\bibinfo {year} {2017})}\BibitemShut {NoStop}%
\bibitem [{\citenamefont {Kamrin}(2010)}]{kamrin2010nonlinear}%
  \BibitemOpen
  \bibfield  {author} {\bibinfo {author} {\bibfnamefont {K.}~\bibnamefont
  {Kamrin}},\ }\bibfield  {title} {\bibinfo {title} {Nonlinear elasto-plastic
  model for dense granular flow},\ }\href@noop {} {\bibfield  {journal}
  {\bibinfo  {journal} {International Journal of Plasticity}\ }\textbf
  {\bibinfo {volume} {26}},\ \bibinfo {pages} {167} (\bibinfo {year}
  {2010})}\BibitemShut {NoStop}%
\bibitem [{\citenamefont {Barker}\ and\ \citenamefont
  {Gray}(2017)}]{barker_gray_2017}%
  \BibitemOpen
  \bibfield  {author} {\bibinfo {author} {\bibfnamefont {T.}~\bibnamefont
  {Barker}}\ and\ \bibinfo {author} {\bibfnamefont {J.~M. N.~T.}\ \bibnamefont
  {Gray}},\ }\bibfield  {title} {\bibinfo {title} {Partial regularisation of
  the incompressible $\mu(\mathrm{I})$-rheology for granular flow},\ }\href
  {https://doi.org/10.1017/jfm.2017.428} {\bibfield  {journal} {\bibinfo
  {journal} {J. Fluid Mech.}\ }\textbf {\bibinfo {volume} {828}},\ \bibinfo
  {pages} {5–32} (\bibinfo {year} {2017})}\BibitemShut {NoStop}%
\bibitem [{\citenamefont {Barker}\ \emph
  {et~al.}(2021{\natexlab{a}})\citenamefont {Barker}, \citenamefont {Rauter},
  \citenamefont {Maguire}, \citenamefont {Johnson},\ and\ \citenamefont
  {Gray}}]{barker2021OpenFoam}%
  \BibitemOpen
  \bibfield  {author} {\bibinfo {author} {\bibfnamefont {T.}~\bibnamefont
  {Barker}}, \bibinfo {author} {\bibfnamefont {M.}~\bibnamefont {Rauter}},
  \bibinfo {author} {\bibfnamefont {E.~S.~F.}\ \bibnamefont {Maguire}},
  \bibinfo {author} {\bibfnamefont {C.~G.}\ \bibnamefont {Johnson}},\ and\
  \bibinfo {author} {\bibfnamefont {J.~M. N.~T.}\ \bibnamefont {Gray}},\
  }\bibfield  {title} {\bibinfo {title} {Coupling rheology and segregation in
  granular flows},\ }\href {https://doi.org/10.1017/jfm.2020.973} {\bibfield
  {journal} {\bibinfo  {journal} {J. Fluid Mech.}\ }\textbf {\bibinfo {volume}
  {909}},\ \bibinfo {pages} {A22} (\bibinfo {year}
  {2021}{\natexlab{a}})}\BibitemShut {NoStop}%
\bibitem [{\citenamefont {Lin}\ and\ \citenamefont
  {Yang}(2020)}]{lin2020continuum}%
  \BibitemOpen
  \bibfield  {author} {\bibinfo {author} {\bibfnamefont {C.-C.}\ \bibnamefont
  {Lin}}\ and\ \bibinfo {author} {\bibfnamefont {F.-L.}\ \bibnamefont {Yang}},\
  }\bibfield  {title} {\bibinfo {title} {Continuum simulation for regularized
  non-local $\mu(\mathrm{I})$ model of dense granular flows},\ }\href@noop {}
  {\bibfield  {journal} {\bibinfo  {journal} {Journal of Computational
  Physics}\ }\textbf {\bibinfo {volume} {420}},\ \bibinfo {pages} {109708}
  (\bibinfo {year} {2020})}\BibitemShut {NoStop}%
\bibitem [{\citenamefont {Chauchat}\ and\ \citenamefont
  {Médale}(2014)}]{chauchat2014three}%
  \BibitemOpen
  \bibfield  {author} {\bibinfo {author} {\bibfnamefont {J.}~\bibnamefont
  {Chauchat}}\ and\ \bibinfo {author} {\bibfnamefont {M.}~\bibnamefont
  {Médale}},\ }\bibfield  {title} {\bibinfo {title} {A three-dimensional
  numerical model for dense granular flows based on the $\mu(\mathrm{I})$
  rheology},\ }\href
  {https://doi.org/https://doi.org/10.1016/j.jcp.2013.09.004} {\bibfield
  {journal} {\bibinfo  {journal} {Journal of Computational Physics}\ }\textbf
  {\bibinfo {volume} {256}},\ \bibinfo {pages} {696} (\bibinfo {year}
  {2014})}\BibitemShut {NoStop}%
\bibitem [{\citenamefont {Henann}\ and\ \citenamefont
  {Kamrin}(2013)}]{henann2013predictive}%
  \BibitemOpen
  \bibfield  {author} {\bibinfo {author} {\bibfnamefont {D.~L.}\ \bibnamefont
  {Henann}}\ and\ \bibinfo {author} {\bibfnamefont {K.}~\bibnamefont
  {Kamrin}},\ }\bibfield  {title} {\bibinfo {title} {A predictive,
  size-dependent continuum model for dense granular flows},\ }\href@noop {}
  {\bibfield  {journal} {\bibinfo  {journal} {Proceedings of the National
  Academy of Sciences}\ }\textbf {\bibinfo {volume} {110}},\ \bibinfo {pages}
  {6730} (\bibinfo {year} {2013})}\BibitemShut {NoStop}%
\bibitem [{\citenamefont {Barker}\ \emph
  {et~al.}(2021{\natexlab{b}})\citenamefont {Barker}, \citenamefont {Rauter},
  \citenamefont {Maguire}, \citenamefont {Johnson},\ and\ \citenamefont
  {Gray}}]{barker_2021_gray}%
  \BibitemOpen
  \bibfield  {author} {\bibinfo {author} {\bibfnamefont {T.}~\bibnamefont
  {Barker}}, \bibinfo {author} {\bibfnamefont {M.}~\bibnamefont {Rauter}},
  \bibinfo {author} {\bibfnamefont {E.~S.~F.}\ \bibnamefont {Maguire}},
  \bibinfo {author} {\bibfnamefont {C.~G.}\ \bibnamefont {Johnson}},\ and\
  \bibinfo {author} {\bibfnamefont {J.~M. N.~T.}\ \bibnamefont {Gray}},\
  }\bibfield  {title} {\bibinfo {title} {Coupling rheology and segregation in
  granular flows},\ }\href {https://doi.org/10.1017/jfm.2020.973} {\bibfield
  {journal} {\bibinfo  {journal} {J. Fluid Mech.}\ }\textbf {\bibinfo {volume}
  {909}},\ \bibinfo {pages} {A22} (\bibinfo {year}
  {2021}{\natexlab{b}})}\BibitemShut {NoStop}%
\bibitem [{\citenamefont {Barker}\ \emph {et~al.}(2015)\citenamefont {Barker},
  \citenamefont {Schaeffer}, \citenamefont {Bohorquez},\ and\ \citenamefont
  {Gray}}]{barker2015well}%
  \BibitemOpen
  \bibfield  {author} {\bibinfo {author} {\bibfnamefont {T.}~\bibnamefont
  {Barker}}, \bibinfo {author} {\bibfnamefont {D.}~\bibnamefont {Schaeffer}},
  \bibinfo {author} {\bibfnamefont {P.}~\bibnamefont {Bohorquez}},\ and\
  \bibinfo {author} {\bibfnamefont {J.}~\bibnamefont {Gray}},\ }\bibfield
  {title} {\bibinfo {title} {Well-posed and ill-posed behaviour of the
  ${\it\mu}(\mathrm{I})$-rheology for granular flow},\ }\href
  {https://doi.org/10.1017/jfm.2015.412} {\bibfield  {journal} {\bibinfo
  {journal} {J. Fluid Mech.}\ }\textbf {\bibinfo {volume} {779}},\ \bibinfo
  {pages} {794–818} (\bibinfo {year} {2015})}\BibitemShut {NoStop}%
\bibitem [{\citenamefont {Dunatunga}\ and\ \citenamefont
  {Kamrin}(2015)}]{dunatunga2015continuum}%
  \BibitemOpen
  \bibfield  {author} {\bibinfo {author} {\bibfnamefont {S.}~\bibnamefont
  {Dunatunga}}\ and\ \bibinfo {author} {\bibfnamefont {K.}~\bibnamefont
  {Kamrin}},\ }\bibfield  {title} {\bibinfo {title} {Continuum modelling and
  simulation of granular flows through their many phases},\ }\href@noop {}
  {\bibfield  {journal} {\bibinfo  {journal} {J. Fluid Mech.}\ }\textbf
  {\bibinfo {volume} {779}},\ \bibinfo {pages} {483} (\bibinfo {year}
  {2015})}\BibitemShut {NoStop}%
\bibitem [{\citenamefont {Franci}\ and\ \citenamefont
  {Cremonesi}(2019)}]{franci20193d}%
  \BibitemOpen
  \bibfield  {author} {\bibinfo {author} {\bibfnamefont {A.}~\bibnamefont
  {Franci}}\ and\ \bibinfo {author} {\bibfnamefont {M.}~\bibnamefont
  {Cremonesi}},\ }\bibfield  {title} {\bibinfo {title} {3$\mathrm{D}$
  regularized $\mu(\mathrm{I})$-rheology for granular flows simulation},\
  }\href@noop {} {\bibfield  {journal} {\bibinfo  {journal} {Journal of
  Computational Physics}\ }\textbf {\bibinfo {volume} {378}},\ \bibinfo {pages}
  {257} (\bibinfo {year} {2019})}\BibitemShut {NoStop}%
\bibitem [{\citenamefont {Parez}\ \emph {et~al.}(2016)\citenamefont {Parez},
  \citenamefont {Aharonov},\ and\ \citenamefont
  {Toussaint}}]{parez2016unsteady}%
  \BibitemOpen
  \bibfield  {author} {\bibinfo {author} {\bibfnamefont {S.}~\bibnamefont
  {Parez}}, \bibinfo {author} {\bibfnamefont {E.}~\bibnamefont {Aharonov}},\
  and\ \bibinfo {author} {\bibfnamefont {R.}~\bibnamefont {Toussaint}},\
  }\bibfield  {title} {\bibinfo {title} {Unsteady granular flows down an
  inclined plane},\ }\href@noop {} {\bibfield  {journal} {\bibinfo  {journal}
  {Phys. Rev. E}\ }\textbf {\bibinfo {volume} {93}},\ \bibinfo {pages} {042902}
  (\bibinfo {year} {2016})}\BibitemShut {NoStop}%
\bibitem [{\citenamefont {Staron}\ \emph {et~al.}(2012)\citenamefont {Staron},
  \citenamefont {Lagrée},\ and\ \citenamefont {Popinet}}]{staron-2012}%
  \BibitemOpen
  \bibfield  {author} {\bibinfo {author} {\bibfnamefont {L.}~\bibnamefont
  {Staron}}, \bibinfo {author} {\bibfnamefont {P.-Y.}\ \bibnamefont
  {Lagrée}},\ and\ \bibinfo {author} {\bibfnamefont {S.}~\bibnamefont
  {Popinet}},\ }\bibfield  {title} {\bibinfo {title} {The granular silo as a
  continuum plastic flow: The hour-glass vs the clepsydra},\ }\href
  {https://doi.org/10.1063/1.4757390} {\bibfield  {journal} {\bibinfo
  {journal} {Phys. Fluids}\ }\textbf {\bibinfo {volume} {24}},\ \bibinfo
  {pages} {103301} (\bibinfo {year} {2012})}\BibitemShut {NoStop}%
\bibitem [{\citenamefont {Staron}\ \emph {et~al.}(2014)\citenamefont {Staron},
  \citenamefont {Lagr{\'e}e},\ and\ \citenamefont
  {Popinet}}]{staron2014continuum}%
  \BibitemOpen
  \bibfield  {author} {\bibinfo {author} {\bibfnamefont {L.}~\bibnamefont
  {Staron}}, \bibinfo {author} {\bibfnamefont {P.-Y.}\ \bibnamefont
  {Lagr{\'e}e}},\ and\ \bibinfo {author} {\bibfnamefont {S.}~\bibnamefont
  {Popinet}},\ }\bibfield  {title} {\bibinfo {title} {Continuum simulation of
  the discharge of the granular silo},\ }\href@noop {} {\bibfield  {journal}
  {\bibinfo  {journal} {The European Physical Journal E}\ }\textbf {\bibinfo
  {volume} {37}},\ \bibinfo {pages} {1} (\bibinfo {year} {2014})}\BibitemShut
  {NoStop}%
\bibitem [{\citenamefont {Rauter}(2021)}]{rauter2021compressible}%
  \BibitemOpen
  \bibfield  {author} {\bibinfo {author} {\bibfnamefont {M.}~\bibnamefont
  {Rauter}},\ }\bibfield  {title} {\bibinfo {title} {The compressible granular
  collapse in a fluid as a continuum: validity of a navier-stokes model with
  $\mu(\mathrm{J}),\phi(\mathrm{J})$-rheology},\ }\href@noop {} {\bibfield
  {journal} {\bibinfo  {journal} {J. Fluid Mech.}\ }\textbf {\bibinfo {volume}
  {915}},\ \bibinfo {pages} {A87} (\bibinfo {year} {2021})}\BibitemShut
  {NoStop}%
\bibitem [{\citenamefont {Xiao}\ \emph {et~al.}(2017)\citenamefont {Xiao},
  \citenamefont {Ottino}, \citenamefont {Lueptow},\ and\ \citenamefont
  {Umbanhowar}}]{boundedheapflow2017}%
  \BibitemOpen
  \bibfield  {author} {\bibinfo {author} {\bibfnamefont {H.}~\bibnamefont
  {Xiao}}, \bibinfo {author} {\bibfnamefont {J.~M.}\ \bibnamefont {Ottino}},
  \bibinfo {author} {\bibfnamefont {R.~M.}\ \bibnamefont {Lueptow}},\ and\
  \bibinfo {author} {\bibfnamefont {P.~B.}\ \bibnamefont {Umbanhowar}},\
  }\bibfield  {title} {\bibinfo {title} {Transient response in granular
  quasi-two-dimensional bounded heap flow},\ }\href
  {https://doi.org/10.1103/PhysRevE.96.040902} {\bibfield  {journal} {\bibinfo
  {journal} {Phys. Rev. E}\ }\textbf {\bibinfo {volume} {96}},\ \bibinfo
  {pages} {040902} (\bibinfo {year} {2017})}\BibitemShut {NoStop}%
\bibitem [{\citenamefont {Pilvar}\ \emph {et~al.}(2019)\citenamefont {Pilvar},
  \citenamefont {Pouraghniaei},\ and\ \citenamefont
  {Shakibaeinia}}]{submergedlandslide2019}%
  \BibitemOpen
  \bibfield  {author} {\bibinfo {author} {\bibfnamefont {M.}~\bibnamefont
  {Pilvar}}, \bibinfo {author} {\bibfnamefont {M.~J.}\ \bibnamefont
  {Pouraghniaei}},\ and\ \bibinfo {author} {\bibfnamefont {A.}~\bibnamefont
  {Shakibaeinia}},\ }\bibfield  {title} {\bibinfo {title} {Two-dimensional
  sub-aerial, submerged, and transitional granular slides},\ }\href
  {https://doi.org/10.1063/1.5121881} {\bibfield  {journal} {\bibinfo
  {journal} {Phys. Fluids}\ }\textbf {\bibinfo {volume} {31}},\ \bibinfo
  {pages} {113303} (\bibinfo {year} {2019})}\BibitemShut {NoStop}%
\bibitem [{\citenamefont {Gray}\ and\ \citenamefont
  {Edwards}(2014)}]{gray2014depth}%
  \BibitemOpen
  \bibfield  {author} {\bibinfo {author} {\bibfnamefont {J.}~\bibnamefont
  {Gray}}\ and\ \bibinfo {author} {\bibfnamefont {A.}~\bibnamefont {Edwards}},\
  }\bibfield  {title} {\bibinfo {title} {A depth-averaged-rheology for shallow
  granular free-surface flows},\ }\href@noop {} {\bibfield  {journal} {\bibinfo
   {journal} {J. Fluid Mech.}\ }\textbf {\bibinfo {volume} {755}},\ \bibinfo
  {pages} {503} (\bibinfo {year} {2014})}\BibitemShut {NoStop}%
\bibitem [{\citenamefont {Rocha}\ \emph {et~al.}(2019)\citenamefont {Rocha},
  \citenamefont {Johnson},\ and\ \citenamefont {Gray}}]{rocha2019self}%
  \BibitemOpen
  \bibfield  {author} {\bibinfo {author} {\bibfnamefont {F.}~\bibnamefont
  {Rocha}}, \bibinfo {author} {\bibfnamefont {C.}~\bibnamefont {Johnson}},\
  and\ \bibinfo {author} {\bibfnamefont {J.}~\bibnamefont {Gray}},\ }\bibfield
  {title} {\bibinfo {title} {Self-channelisation and levee formation in
  monodisperse granular flows},\ }\href@noop {} {\bibfield  {journal} {\bibinfo
   {journal} {J. Fluid Mech.}\ }\textbf {\bibinfo {volume} {876}},\ \bibinfo
  {pages} {591} (\bibinfo {year} {2019})}\BibitemShut {NoStop}%
\bibitem [{\citenamefont {Delannay}\ \emph {et~al.}(2017)\citenamefont
  {Delannay}, \citenamefont {Valance}, \citenamefont {Mangeney}, \citenamefont
  {Roche},\ and\ \citenamefont {Richard}}]{delannay2017granular}%
  \BibitemOpen
  \bibfield  {author} {\bibinfo {author} {\bibfnamefont {R.}~\bibnamefont
  {Delannay}}, \bibinfo {author} {\bibfnamefont {A.}~\bibnamefont {Valance}},
  \bibinfo {author} {\bibfnamefont {A.}~\bibnamefont {Mangeney}}, \bibinfo
  {author} {\bibfnamefont {O.}~\bibnamefont {Roche}},\ and\ \bibinfo {author}
  {\bibfnamefont {P.}~\bibnamefont {Richard}},\ }\bibfield  {title} {\bibinfo
  {title} {Granular and particle-laden flows: from laboratory experiments to
  field observations},\ }\href@noop {} {\bibfield  {journal} {\bibinfo
  {journal} {Journal of Physics D: Applied Physics}\ }\textbf {\bibinfo
  {volume} {50}},\ \bibinfo {pages} {053001} (\bibinfo {year}
  {2017})}\BibitemShut {NoStop}%
\bibitem [{\citenamefont {Mangeney}\ \emph {et~al.}(2007)\citenamefont
  {Mangeney}, \citenamefont {Bouchut}, \citenamefont {Thomas}, \citenamefont
  {Vilotte},\ and\ \citenamefont {Bristeau}}]{mangeney2007numerical}%
  \BibitemOpen
  \bibfield  {author} {\bibinfo {author} {\bibfnamefont {A.}~\bibnamefont
  {Mangeney}}, \bibinfo {author} {\bibfnamefont {F.}~\bibnamefont {Bouchut}},
  \bibinfo {author} {\bibfnamefont {N.}~\bibnamefont {Thomas}}, \bibinfo
  {author} {\bibfnamefont {J.-P.}\ \bibnamefont {Vilotte}},\ and\ \bibinfo
  {author} {\bibfnamefont {M.-O.}\ \bibnamefont {Bristeau}},\ }\bibfield
  {title} {\bibinfo {title} {Numerical modeling of self-channeling granular
  flows and of their levee-channel deposits},\ }\href@noop {} {\bibfield
  {journal} {\bibinfo  {journal} {Journal of Geophysical Research: Earth
  Surface}\ }\textbf {\bibinfo {volume} {112}} (\bibinfo {year}
  {2007})}\BibitemShut {NoStop}%
\bibitem [{\citenamefont {Dsouza}\ and\ \citenamefont
  {Nott}(2020)}]{dsouzaNott2020nonlocal}%
  \BibitemOpen
  \bibfield  {author} {\bibinfo {author} {\bibfnamefont {P.~V.}\ \bibnamefont
  {Dsouza}}\ and\ \bibinfo {author} {\bibfnamefont {P.~R.}\ \bibnamefont
  {Nott}},\ }\bibfield  {title} {\bibinfo {title} {A non-local constitutive
  model for slow granular flow that incorporates dilatancy},\ }\href@noop {}
  {\bibfield  {journal} {\bibinfo  {journal} {J. Fluid Mech.}\ }\textbf
  {\bibinfo {volume} {888}} (\bibinfo {year} {2020})}\BibitemShut {NoStop}%
\bibitem [{\citenamefont {Pouliquen}\ and\ \citenamefont
  {Forterre}(2009)}]{pouliquen2009non}%
  \BibitemOpen
  \bibfield  {author} {\bibinfo {author} {\bibfnamefont {O.}~\bibnamefont
  {Pouliquen}}\ and\ \bibinfo {author} {\bibfnamefont {Y.}~\bibnamefont
  {Forterre}},\ }\bibfield  {title} {\bibinfo {title} {A non-local rheology for
  dense granular flows},\ }\href@noop {} {\bibfield  {journal} {\bibinfo
  {journal} {Philosophical Transactions of the Royal Society A: Mathematical,
  Physical and Engineering Sciences}\ }\textbf {\bibinfo {volume} {367}},\
  \bibinfo {pages} {5091} (\bibinfo {year} {2009})}\BibitemShut {NoStop}%
\bibitem [{\citenamefont {Kamrin}\ and\ \citenamefont
  {Koval}(2012)}]{kamrin2012nonlocal}%
  \BibitemOpen
  \bibfield  {author} {\bibinfo {author} {\bibfnamefont {K.}~\bibnamefont
  {Kamrin}}\ and\ \bibinfo {author} {\bibfnamefont {G.}~\bibnamefont {Koval}},\
  }\bibfield  {title} {\bibinfo {title} {Nonlocal constitutive relation for
  steady granular flow},\ }\href@noop {} {\bibfield  {journal} {\bibinfo
  {journal} {Phys. Rev. Lett.}\ }\textbf {\bibinfo {volume} {108}},\ \bibinfo
  {pages} {178301} (\bibinfo {year} {2012})}\BibitemShut {NoStop}%
\bibitem [{\citenamefont {Debnath}\ \emph
  {et~al.}(2022{\natexlab{b}})\citenamefont {Debnath}, \citenamefont
  {Kumaran},\ and\ \citenamefont {Rao}}]{debnath2023comparison}%
  \BibitemOpen
  \bibfield  {author} {\bibinfo {author} {\bibfnamefont {B.}~\bibnamefont
  {Debnath}}, \bibinfo {author} {\bibfnamefont {V.}~\bibnamefont {Kumaran}},\
  and\ \bibinfo {author} {\bibfnamefont {K.~K.}\ \bibnamefont {Rao}},\
  }\bibfield  {title} {\bibinfo {title} {Comparison of the compressible
  $\mu(\mathrm{I})$ class of models and non-local models with the discrete
  element method for steady fully developed flow of cohesionless granular
  materials through a vertical channel},\ }\href
  {https://doi.org/10.1017/jfm.2022.119} {\bibfield  {journal} {\bibinfo
  {journal} {J. Fluid Mech.}\ }\textbf {\bibinfo {volume} {937}},\ \bibinfo
  {pages} {A33} (\bibinfo {year} {2022}{\natexlab{b}})}\BibitemShut {NoStop}%
\bibitem [{\citenamefont {Barker}\ \emph {et~al.}(2017)\citenamefont {Barker},
  \citenamefont {Schaeffer}, \citenamefont {Shearer},\ and\ \citenamefont
  {Gray}}]{barker2017wellposedrheology}%
  \BibitemOpen
  \bibfield  {author} {\bibinfo {author} {\bibfnamefont {T.}~\bibnamefont
  {Barker}}, \bibinfo {author} {\bibfnamefont {D.~G.}\ \bibnamefont
  {Schaeffer}}, \bibinfo {author} {\bibfnamefont {M.}~\bibnamefont {Shearer}},\
  and\ \bibinfo {author} {\bibfnamefont {J.~M. N.~T.}\ \bibnamefont {Gray}},\
  }\bibfield  {title} {\bibinfo {title} {Well-posed continuum equations for
  granular flow with compressibility and $\mu(\mathrm{I})$-rheology},\
  }\href@noop {} {\bibfield  {journal} {\bibinfo  {journal} {Proceedings of the
  Royal Society A: Mathematical, Physical and Engineering Sciences}\ }\textbf
  {\bibinfo {volume} {473}},\ \bibinfo {pages} {20160846} (\bibinfo {year}
  {2017})}\BibitemShut {NoStop}%
\bibitem [{\citenamefont {Schaeffer}\ \emph {et~al.}(2019)\citenamefont
  {Schaeffer}, \citenamefont {Barker}, \citenamefont {Tsuji}, \citenamefont
  {Gremaud}, \citenamefont {Shearer},\ and\ \citenamefont
  {Gray}}]{schaeffer2019constitutive}%
  \BibitemOpen
  \bibfield  {author} {\bibinfo {author} {\bibfnamefont {D.}~\bibnamefont
  {Schaeffer}}, \bibinfo {author} {\bibfnamefont {T.}~\bibnamefont {Barker}},
  \bibinfo {author} {\bibfnamefont {D.}~\bibnamefont {Tsuji}}, \bibinfo
  {author} {\bibfnamefont {P.}~\bibnamefont {Gremaud}}, \bibinfo {author}
  {\bibfnamefont {M.}~\bibnamefont {Shearer}},\ and\ \bibinfo {author}
  {\bibfnamefont {J.}~\bibnamefont {Gray}},\ }\bibfield  {title} {\bibinfo
  {title} {Constitutive relations for compressible granular flow in the
  inertial regime},\ }\href@noop {} {\bibfield  {journal} {\bibinfo  {journal}
  {J. Fluid Mech.}\ }\textbf {\bibinfo {volume} {874}},\ \bibinfo {pages} {926}
  (\bibinfo {year} {2019})}\BibitemShut {NoStop}%
\bibitem [{\citenamefont {Kamrin}(2019)}]{kamrin2019non}%
  \BibitemOpen
  \bibfield  {author} {\bibinfo {author} {\bibfnamefont {K.}~\bibnamefont
  {Kamrin}},\ }\bibfield  {title} {\bibinfo {title} {Non-locality in granular
  flow: Phenomenology and modeling approaches},\ }\href@noop {} {\bibfield
  {journal} {\bibinfo  {journal} {Frontiers in Physics}\ }\textbf {\bibinfo
  {volume} {7}},\ \bibinfo {pages} {116} (\bibinfo {year} {2019})}\BibitemShut
  {NoStop}%
\bibitem [{\citenamefont {Heyman}\ \emph
  {et~al.}(2017{\natexlab{b}})\citenamefont {Heyman}, \citenamefont {Delannay},
  \citenamefont {Tabuteau},\ and\ \citenamefont
  {Valance}}]{heyman2017compressibility}%
  \BibitemOpen
  \bibfield  {author} {\bibinfo {author} {\bibfnamefont {J.}~\bibnamefont
  {Heyman}}, \bibinfo {author} {\bibfnamefont {R.}~\bibnamefont {Delannay}},
  \bibinfo {author} {\bibfnamefont {H.}~\bibnamefont {Tabuteau}},\ and\
  \bibinfo {author} {\bibfnamefont {A.}~\bibnamefont {Valance}},\ }\bibfield
  {title} {\bibinfo {title} {Compressibility regularizes the
  $\mu(\mathrm{I})$-rheology for dense granular flows},\ }\href@noop {}
  {\bibfield  {journal} {\bibinfo  {journal} {J. Fluid Mech.}\ }\textbf
  {\bibinfo {volume} {830}},\ \bibinfo {pages} {553} (\bibinfo {year}
  {2017}{\natexlab{b}})}\BibitemShut {NoStop}%
\bibitem [{\citenamefont {Goldhirsch}\ and\ \citenamefont
  {Sela}(1996)}]{goldhirsch1996origin}%
  \BibitemOpen
  \bibfield  {author} {\bibinfo {author} {\bibfnamefont {I.}~\bibnamefont
  {Goldhirsch}}\ and\ \bibinfo {author} {\bibfnamefont {N.}~\bibnamefont
  {Sela}},\ }\bibfield  {title} {\bibinfo {title} {Origin of normal stress
  differences in rapid granular flows},\ }\href@noop {} {\bibfield  {journal}
  {\bibinfo  {journal} {Phys. Rev. E}\ }\textbf {\bibinfo {volume} {54}},\
  \bibinfo {pages} {4458} (\bibinfo {year} {1996})}\BibitemShut {NoStop}%
\bibitem [{\citenamefont {Alam}\ and\ \citenamefont
  {Luding}(2003)}]{alam2003first}%
  \BibitemOpen
  \bibfield  {author} {\bibinfo {author} {\bibfnamefont {M.}~\bibnamefont
  {Alam}}\ and\ \bibinfo {author} {\bibfnamefont {S.}~\bibnamefont {Luding}},\
  }\bibfield  {title} {\bibinfo {title} {First normal stress difference and
  crystallization in a dense sheared granular fluid},\ }\href@noop {}
  {\bibfield  {journal} {\bibinfo  {journal} {Phys. Fluids}\ }\textbf {\bibinfo
  {volume} {15}},\ \bibinfo {pages} {2298} (\bibinfo {year}
  {2003})}\BibitemShut {NoStop}%
\bibitem [{\citenamefont {Saha}\ and\ \citenamefont
  {Alam}(2016)}]{saha2016normal}%
  \BibitemOpen
  \bibfield  {author} {\bibinfo {author} {\bibfnamefont {S.}~\bibnamefont
  {Saha}}\ and\ \bibinfo {author} {\bibfnamefont {M.}~\bibnamefont {Alam}},\
  }\bibfield  {title} {\bibinfo {title} {Normal stress differences, their
  origin and constitutive relations for a sheared granular fluid},\ }\href@noop
  {} {\bibfield  {journal} {\bibinfo  {journal} {J. Fluid Mech.}\ }\textbf
  {\bibinfo {volume} {795}},\ \bibinfo {pages} {549} (\bibinfo {year}
  {2016})}\BibitemShut {NoStop}%
\bibitem [{\citenamefont {Srivastava}\ \emph {et~al.}(2021)\citenamefont
  {Srivastava}, \citenamefont {Silbert}, \citenamefont {Grest},\ and\
  \citenamefont {Lechman}}]{srivastava2021viscometric}%
  \BibitemOpen
  \bibfield  {author} {\bibinfo {author} {\bibfnamefont {I.}~\bibnamefont
  {Srivastava}}, \bibinfo {author} {\bibfnamefont {L.~E.}\ \bibnamefont
  {Silbert}}, \bibinfo {author} {\bibfnamefont {G.~S.}\ \bibnamefont {Grest}},\
  and\ \bibinfo {author} {\bibfnamefont {J.~B.}\ \bibnamefont {Lechman}},\
  }\bibfield  {title} {\bibinfo {title} {Viscometric flow of dense granular
  materials under controlled pressure and shear stress},\ }\href@noop {}
  {\bibfield  {journal} {\bibinfo  {journal} {J. Fluid Mech.}\ }\textbf
  {\bibinfo {volume} {907}},\ \bibinfo {pages} {A18} (\bibinfo {year}
  {2021})}\BibitemShut {NoStop}%
\bibitem [{\citenamefont {Santos}\ \emph {et~al.}(2022)\citenamefont {Santos},
  \citenamefont {Srivastava}, \citenamefont {Silbert}, \citenamefont
  {Lechman},\ and\ \citenamefont {Grest}}]{santos2022}%
  \BibitemOpen
  \bibfield  {author} {\bibinfo {author} {\bibfnamefont {A.~P.}\ \bibnamefont
  {Santos}}, \bibinfo {author} {\bibfnamefont {I.}~\bibnamefont {Srivastava}},
  \bibinfo {author} {\bibfnamefont {L.~E.}\ \bibnamefont {Silbert}}, \bibinfo
  {author} {\bibfnamefont {J.~B.}\ \bibnamefont {Lechman}},\ and\ \bibinfo
  {author} {\bibfnamefont {G.~S.}\ \bibnamefont {Grest}},\ }\bibfield  {title}
  {\bibinfo {title} {Fluctuations and power-law scaling of dry, frictionless
  granular rheology near the hard-particle limit},\ }\href
  {https://doi.org/10.1103/PhysRevFluids.7.084303} {\bibfield  {journal}
  {\bibinfo  {journal} {Phys. Rev. Fluids}\ }\textbf {\bibinfo {volume} {7}},\
  \bibinfo {pages} {084303} (\bibinfo {year} {2022})}\BibitemShut {NoStop}%
\bibitem [{\citenamefont {Shojaaee}\ \emph {et~al.}(2012)\citenamefont
  {Shojaaee}, \citenamefont {Roux}, \citenamefont {Chevoir},\ and\
  \citenamefont {Wolf}}]{shearflowofgranularmaterials2012}%
  \BibitemOpen
  \bibfield  {author} {\bibinfo {author} {\bibfnamefont {Z.}~\bibnamefont
  {Shojaaee}}, \bibinfo {author} {\bibfnamefont {J.-N.}\ \bibnamefont {Roux}},
  \bibinfo {author} {\bibfnamefont {F.}~\bibnamefont {Chevoir}},\ and\ \bibinfo
  {author} {\bibfnamefont {D.~E.}\ \bibnamefont {Wolf}},\ }\bibfield  {title}
  {\bibinfo {title} {Shear flow of dense granular materials near smooth walls.
  $\mathrm{I}$. shear localization and constitutive laws in the boundary
  region},\ }\href {https://doi.org/10.1103/PhysRevE.86.011301} {\bibfield
  {journal} {\bibinfo  {journal} {Phys. Rev. E}\ }\textbf {\bibinfo {volume}
  {86}},\ \bibinfo {pages} {011301} (\bibinfo {year} {2012})}\BibitemShut
  {NoStop}%
\bibitem [{\citenamefont {{Forterre, Yoel and Pouliquen,
  Olivier}}(2002)}]{forterre_instability_2002}%
  \BibitemOpen
  \bibfield  {author} {\bibinfo {author} {\bibnamefont {{Forterre, Yoel and
  Pouliquen, Olivier}}},\ }\bibfield  {title} {\bibinfo {title} {Stability
  analysis of rapid granular chute flows: formation of longitudinal vortices},\
  }\href {https://doi.org/10.1017/S0022112002001581} {\bibfield  {journal}
  {\bibinfo  {journal} {J. Fluid Mech.}\ }\textbf {\bibinfo {volume} {467}},\
  \bibinfo {pages} {361–387} (\bibinfo {year} {2002})}\BibitemShut {NoStop}%
\bibitem [{\citenamefont {R.~B.~Bird}\ and\ \citenamefont
  {Lightfoot}(2007)}]{BSL_2007}%
  \BibitemOpen
  \bibfield  {author} {\bibinfo {author} {\bibfnamefont {W.~E.~S.}\
  \bibnamefont {R.~B.~Bird}}\ and\ \bibinfo {author} {\bibfnamefont {E.~N.}\
  \bibnamefont {Lightfoot}},\ }\href@noop {} {\emph {\bibinfo {title}
  {Transport Phenomena, 2nd Edition}}}\ (\bibinfo  {publisher} {John Wiley \&
  Sons},\ \bibinfo {address} {New York},\ \bibinfo {year} {2007})\BibitemShut
  {NoStop}%
\bibitem [{\citenamefont {Louge}\ \emph {et~al.}(2005)\citenamefont {Louge},
  \citenamefont {Valance}, \citenamefont {Taberlet}, \citenamefont {Richard},\
  and\ \citenamefont {Delannay}}]{louge2005volume}%
  \BibitemOpen
  \bibfield  {author} {\bibinfo {author} {\bibfnamefont {M.}~\bibnamefont
  {Louge}}, \bibinfo {author} {\bibfnamefont {A.}~\bibnamefont {Valance}},
  \bibinfo {author} {\bibfnamefont {N.}~\bibnamefont {Taberlet}}, \bibinfo
  {author} {\bibfnamefont {P.}~\bibnamefont {Richard}},\ and\ \bibinfo {author}
  {\bibfnamefont {R.}~\bibnamefont {Delannay}},\ }\bibfield  {title} {\bibinfo
  {title} {Volume fraction profile in channeled granular flows down an erodible
  incline},\ }\href@noop {} {\bibfield  {journal} {\bibinfo  {journal} {Proc.
  Powders and Grains}\ } (\bibinfo {year} {2005})}\BibitemShut {NoStop}%
\end{thebibliography}%
\end{document}